\documentclass[usenatbib]{mn2e}

\usepackage{times}
\usepackage{lscape}
\usepackage{graphicx}
\usepackage{float}
\usepackage{subfigure}
\usepackage{url}
\usepackage{rotating}

\author[S. Alaghband-Zadeh et al.]
{S.\ Alaghband-Zadeh,$^1$ S.\ C.\ Chapman,$^{1,2}$ A.\ M.\ Swinbank,$^3$ Ian Smail,$^3$ A.\ L.\ R.\ Danielson,$^4$
\newauthor
R.\ Decarli,$^5$ R.\ J.\ Ivison,$^{6,7}$ R. Meijerink,$^8$ A.\ Weiss,$^9$ Paul.\ P.\ van der Werf$^{10}$\\
$^{1}${Institute of Astronomy, Madingley Road, Cambridge, CB3 0HA UK}\\
$^{2}${Department of Physics and Atmospheric Science, Dalhousie University, 6310 Coburg Road, Halifax, Nova Scotia, B3H 4R2, Canada}\\
$^{3}${Institute for Computational Cosmology, Durham University, South Rd, Durham DH1 3LE, UK}\\
$^{4}${Department of Physics, Durham University, South Rd, Durham DH1 3LE, UK}\\
$^{5}${Max-Planck Institut für Astronomie, Königstuhl 17, D-69117 Heidelberg, Germany}\\
$^{6}${UK Astronomy Technology Centre, Science and Technology Facilities Council, Royal Observatory, Blackford Hill, Edinburgh EH9 3HJ}\\
$^{7}${Institute for Astronomy, University of Edinburgh, Blackford Hill, Edinburgh EH9 3HJ}\\
$^{8}${University of Groningen, Kapteyn Astronomical Institute, Landleven 12, 9747 AD Groningen, The Netherlands}\\
$^{9}${Max-Planck Institut für Radioastronomie, Auf dem Hügel 69, 53121 Bonn, Germany}\\
$^{10}${Leiden Observatory, Leiden University, P.O. Box 9513, NL-2300 RA Leiden, The Netherlands}\\
}

\date{\today}
\begin{document}

\title[Using {[}CI{]} to probe the Interstellar Medium in $z\sim2.5$ SMGs] {Using [C{\sc i}] to probe the Interstellar Medium in $z\sim2.5$ Sub-Millimeter Galaxies \footnotemark}


\maketitle
\begin{abstract}
We present new [C{\sc i}](1--0) and $^{12}$CO(4--3)  Plateau de Bure Interferometer (PdBI) observations of five Sub-Millimeter Galaxies (SMGs) and combine these with all available [C{\sc i}](1--0) literature detections in SMGs to probe the gas distribution within a sample of 14 systems. We explore the [C{\sc i}](1--0) properties of the SMG population, particularly investigating the ratio of the [C{\sc i}](1--0) luminosity to various $^{12}$CO transition and far-infrared luminosities. We find that the SMGs with new observations extend the spread of $\rm L_{[CI](1-0)}/L_{FIR}$ to much higher values than found before, with our complete sample providing a good representation of the diverse $z>2$ SMG population.  We compare the line ratios to the outputs of photodissociation region (PDR) models to constrain the physical conditions in the interstellar medium (ISM) of the SMGs, finding an average density of $\langle \log (n/\rm cm^{-3})\rangle=4.3\pm0.2$ and an average radiation field (in terms of the local field value, $G_0$) of $\langle \log (G_0)\rangle=3.9\pm0.4$. Overall, we find the SMGs are most comparable to local ULIRGs in $G_0$ and $n$, however a significant tail of 5 of the 14 SMGs are likely best compared to less compact, local starburst galaxies, providing new evidence that many SMGs have extended star formation distributions and are therefore not simply scaled up versions of local ULIRGs. We derive the ISM properties of a sample of quasars also finding that they have higher densities and radiation fields on average than the SMGs, consistent with the more extreme local ULIRGs, and reinforcing their interpretation as transition objects. We explore the limitations of using simple PDR models to understand [C{\sc i}], which may be concomitant with the bulk  H$_2$ mass rather than PDR-distributed. We therefore also assess [C{\sc i}] as a tracer of H$_2$, finding that for our sample SMGs, the H$_2$ masses derived from [C{\sc i}] are often consistent with those determined from low excitation $^{12}$CO. We conclude that [C{\sc i}] observations provide a useful tool to probe the bulk gas and gas processes occurring within merging SMGs, however more detailed, resolved observations are required to fully exploit [C{\sc i}] as a diagnostic.
\end{abstract}

\begin{keywords}
galaxies: evolution - galaxies: high redshift - galaxies: starburst - galaxies: ISM
\end{keywords}

\footnotetext{Based on observations carried out with the IRAM Plateau de Bure Interferometer. IRAM is supported by INSU/CNRS (France), MPG (Germany) and IGN (Spain).}

\section{Introduction}
\label{sec:intro}
The ultra-luminous Sub-Millimetre Galaxies (SMGs) contribute significantly to the rapid build-up of stellar mass in the Universe at $z\sim2$ (e.g., \citealt{Chapman05}). They have large molecular gas reservoirs with gas masses of $\sim$10$^{10}$\,M$_{\odot}$ \citep{Greve05,Bothwell12} and high star formation rates (SFRs) implying that they can build a stellar mass of 10$^{10}$\,M$_{\odot}$ in only 100\,Myr of continuous star formation  \citep{Magnelli12}, and are therefore often interpreted as being the progenitors of the massive ellipticals observed locally.  Investigating the dense cool interstellar medium (ISM) in high-redshift SMGs is important for understanding the mechanisms for triggering these high SFRs and probing the fuel supply for the star formation.

The immense luminosities in local ULIRGs often appear to be triggered by mergers and interactions \citep{Sanders96}. However the star formation in local ULIRGs is much more centrally concentrated than high-$z$ ULIRGs \citep{Sakamoto08} where carbon monoxide ($^{12}$CO) and radio observations have shown that the gas can be extended on scales up to $\sim10$\,kpc and often comprise multiple components \citep{Chapman04b, Biggs08, Tacconi08, Bothwell10, Ivison11} highlighting a difference in the local and high-redshift populations. \cite{Menendez-Delmestre09} also suggest the SMGs contain extended cool and warm dust distributions from mid-infrared spectroscopic observations of the polycyclic aromatic hydrocarbon (PAH) emission. Observations of the [C{\sc ii}] cooling line in SMM J213511, the `Eyelash', suggest that this system consists of extended starburst clumps, providing further support that SMGs are not scaled up versions of local ULIRGs \citep{Ivison10b} whilst studies of the rest-frame near-infrared colours of SMGs indicate that the SMGs have more widely distributed star formation than the compact local ULIRGs \citep{Hainline09}. Studies of carbon monoxide transitions and the cooling lines in the SMGs can therefore be used to search for the trigger of the intense star formation by determining the extent of the fuel for the star formation. 

Studying various $^{12}$CO transitions can help constrain the physical conditions within the galaxies. However with only $^{12}$CO measurements, a number of properties of the molecular gas are degenerate (density and kinetic temperature; \citealt{Walter11}) so other tracers are required to break this degeneracy, for example multiple $^{13}$CO lines, HCN and [C{\sc i}].  The ratios of the [C{\sc i}] lines with other $^{12}$CO transitions have been suggested as diagnostics in photodissociation region (PDR) models in which far-ultraviolet photons illuminate molecular clouds \citep{Hollenbach99} controlling the properties of the interstellar medium (ISM). In the one dimensional PDR slab models of \cite{Kaufman99} a relatively thin layer of neutral atomic gas is predicted in between the transition from the C+ rich outer layer to the CO rich inner volume,  naively implying that the [C{\sc i}] should not be able to trace the bulk H$_2$ mass. However there is observational evidence that [C{\sc i}] does trace the same distribution as the $^{12}$CO and therefore the  H$_2$, as discussed by \cite{Papadopoulos04a}. The two  [C{\sc i}] transitions are easily excited and provide a more complete H$_2$ mass tracer than the high-J $^{12}$CO lines found closest in frequency ($^{12}$CO(4--3) and $^{12}$CO(7--6)), since high excitation $^{12}$CO is unequivocally tied to the star-forming, warm and dense gas. Viable explanations for [C{\sc i}] spatially correlating to the bulk H$_2$ mass tracers employ clumpy, inhomogeneous PDR models \citep{Meixner93,Spaans96} where the surface layers of [C{\sc i}] are distributed across the clouds in clumps. The widespread  [C{\sc i}] distribution is therefore predicted in the PDR model of \cite{Spaans96} however the intensity correlation has yet to be tested over a wide range of environments \citep{Keene96}.

Observations of  [C{\sc i}]  transitions have shown that the ratio of L$'_{\rm [CI](^3P_2\rightarrow\,^3P_1)}/$\,L$'_{\rm  [CI](^3P_1\rightarrow\,^3P_0)}$ (where L$'$ is the line luminosity in units of K\,kms$^{-1}$\,pc$^2$) provides a sensitive probe of the temperature of the interstellar medium at moderate densities.  Both lines have modest critical densities ($n_{\rm crit}\sim 0.3$--$1.1\times10^3$\,cm$^{-3}$) and are therefore often thermalised in molecular clouds with $n> 10^{3}$\,cm$^{-3}$.  The lines arise from states with energy levels T$_1=23.6$\,K and T$_2 =62.5$\,K above the ground state, and thus their ratio is sensitive to the gas temperature if T$_{\rm gas}<100$\,K. Furthermore,  the H$_2$ gas mass determined from the [C{\sc i}] line agrees with the H$_2$ mass from multi-transition $^{12}$CO and dust continuum observations in local ULIRGs \citep{Papadopoulos04b}, Arp220 \citep{Gerin98} and the Cloverleaf QSO \citep{Weiss03}, validating its utility across a large range in physical environments.

Since its initial detection in interstellar clouds  \citep{Phillips81}, [C{\sc i}] has been detected in a number of local systems, including near the Galactic centre \citep{Ojha01}, in local giant molecular clouds (Orion A and B; \citealt{Ikeda02}), nearby external galaxies \citep{Israel02, Gerin02} and in local ULIRGs \citep{Papadopoulos04b}. However, [C{\sc i}] has only been detected in a small number of high-redshift systems, mainly gravitationally lensed SMGs and quasars, some of which  may suffer from differential lensing uncertainties between potentially vast reservoirs of low-excitation gas and dense regions of star forming gas  \citep{Wagg06, Weiss05,Ao08, Riechers09, Barvainis97, Lestrade10, Danielson11,Pety04}. 

Initial studies of [C{\sc i}] in SMGs have been carried out by \cite{Walter11} who present detections of [C{\sc i}](1--0) and [C{\sc i}](2--1) in a sample of 14 SMGs and quasars, concluding that the [C{\sc i}] properties of these high-redshift systems are not significantly different to local systems and the Milky Way. However, the majority of the \cite{Walter11} SMG sample are either strongly gravitationally lensed or have strong AGN components, or both. There is, therefore, a need to understand the [C{\sc i}] gas distribution in a sample of un-lensed SMGs whose properties are dominated by the star formation rather than an AGN. In this paper we study the [C{\sc i}] line in a sample of SMGs, five of which have  newly obtained observations with the remainder taken from literature, to test its use as a diagnostic of the ISM physical conditions within SMGs.  We explore the dichotomy in the use of  [C{\sc i}] as a tracer of the bulk H$_2$ mass  and its use in assessing the outputs from PDR modelling relative to well studied sources and a range of galaxy types. All calculations assume a flat, $\Lambda$ cold dark matter ($\Lambda$CDM) cosmology with $\Omega_\Lambda=0.7$ and $H_0=71$\,kms$^{-1}$Mpc$^{-1}$.

\section{Sample, Observations and Reduction} 
\label{sec:obs}
To complement the initial studies of  [C{\sc i}] in SMGs from \cite{Walter11}, we selected five additional, un-lensed, $z\sim2.3$  SMGs from extensive spectroscopic surveys expanding the \cite{Chapman05} sample. Two of these SMGs are components in a wide-separation major merger, as presented in \cite{Alaghband-Zadeh12}. The SMGs were observed in [C{\sc i}]($^3P_1\rightarrow\,^3P_0$) (rest frequency: 492.161\,GHz), which we refer to as [C{\sc i}](1--0) from here on, and $^{12}$CO(4--3) (rest frequency: 461.041\,GHz). These SMGs were chosen to have well detected, resolved H$\alpha$ lines to enable direct comparison of the ionized gas component in these galaxies. They were also all selected to lie in equatorial fields to enable further followup with the Atacama Large Millimeter/submillimeter Array (ALMA). The redshift range we covered matches the peak of the SMG population ($z=2-2.5$ -- \citealt{Chapman05, Wardlow11}).  These sources have been mapped in H$\alpha$ from integral field spectroscopy observations made with Gemini-NIFS and VLT-SINFONI. The H$\alpha$ intensity maps show multiple peaks of star formation and the velocity and velocity dispersion maps are turbulent and disturbed, distinguishing them from smooth, disk-like systems. These SMGs are classified as mergers using a variety of criteria, including a kinemetry analysis (\citealt{Alaghband-Zadeh12}). Four of our new sources do not show any signs of AGN in any wavelength, while the fifth has a broad AGN (a component in the large separation ($\sim20$kpc)  major merger). The combined sample these new targets and the \cite{Walter11} SMGs displays a large range of environments (different stages of mergers and a range of AGN contributions) and are well suited to assessing the diagnostic power of the [C{\sc i}] line.

\begin{table*}
\centering
\small
\begin{tabular}{|l|l|c|c|c|c|c|c|c|c|c|}
\hline\hline
ID &   RA & Dec & z & t$\rm_{CO(4-3)}$  & t$\rm_{[CI](1-0)}$  & $\rm S_{\rm 2mm}$  & $\rm S_{\rm 850\mu m}$ & $\rm \nu_{obs,CO(4-3)}$ & $\rm \nu_{obs, [CI](1-0)}$\\
 {} & {} & {} & {} & (hrs) & (hrs) & (mJy) & (mJy) & GHz & GHz \\
\hline\hline
SXDF7 {\it (J0217-0505)}   &  02:17:38.92 & -05:05:23.7  & 2.5286[8] & 3.2 & 5.1  & 0.3$\pm$0.1 & 7.1$\pm$1.5 & 130.6  & 139.4\\    
SXDF11 {\it (J0217-0459)} &  02:17:25.12 & -04:59:37.4  & 2.2821[8]  & 3.3 & 3.3  & -           & 4.5$\pm$1.9 & 138.5  & 148.9\\ 
SXDF4a {\it (J0217-0503a)} &  02:17:38.62 & -05:03:37.5  & 2.0298[2] & 2.1 & 7.8   & 0.5$\pm$0.1 & 2.2$\pm$0.85 & 151.9& 162.2 \\
SXDF4b {\it (J0217-0503b)} &  02:17:38.62 & -05:03:37.5  & 2.0274[1] & 2.1 & 7.8   & 0.4$\pm$0.1 & 2.2$\pm$0.85 &151.9 &  162.2\\
SA22.96 {\it (J2218-0021)} &  22:18:04.42 &  00:21:54.4  & 2.5169[7]  & 3.0 & 5.3  & 0.2$\pm$0.1 & 9.0$\pm$2.3 &131.1 & 139.9\\ 
\hline\hline
\end{tabular}
\caption{SMG positions from the 1.4\,GHz detections, and redshifts from the second order moment analysis of the $^{12}$CO(4--3) observations of this work. The IDs in italics indicate the IDs used for the H$\alpha$ observations in \protect\cite{Alaghband-Zadeh12}. The 2\,mm continuum is measured in the off-line regions of the [C{\sc i}](1--0) observations, which are deeper and also at slightly higher frequency than the $^{12}$CO(4--3) observations. The 850\,$\mu$m fluxes are from \protect\cite{Chapman05} and \protect\cite{Coppin06}. The values in brackets in the redshift column represent the error on the last decimal place. Since we detect approximately equal 2\,mm continuum across both SXDF4a and SXDF4b, the $S_{\rm 850\mu m}$ values for SXDF4a and SXDF4b assume the measured 850\,$\mu$m flux for the whole system is equally divided between the two components. We also quote the 5 antennae on-source times (t) and the observing frequencies ($\nu$).}
\label{tab:sample}
\end{table*}

These additional SMGs were observed with the Plateau de Bure Interferometer (PdBI) in 2011 in D configuration.  The summer observations used a  sub-array, 5 antennae configuration, while the 6th antenna was being serviced. The  SMGs were observed at their redshifted [C{\sc i}] and $^{12}$CO(4--3) frequencies lying well within the band of the 2mm receivers (Table \ref{tab:sample}).

The observations were reduced using the {\sc GILDAS} software from IRAM. Data were flagged and removed that had high phase RMS, as well as other data with large amplitude losses, pointing, focus or tracking errors. The flux calibrators used for the $^{12}$CO(4--3) observations were a selection of 3C454, 3C454.3, MWC349 and 1749-096 and the phase and amplitude calibrators were a selection of 0130-171, 0138-097, 0237-027 and 2223-052. The flux calibrators used for the [C{\sc i}](1--0) observations were a selection of 3C454.3, MWC349 and 1749-096 and the phase and amplitude calibrators were a selection of 0130-171, 0106+013 and 2223-052. The seeing ranged from 0.91$''$ to 2.02$''$. The phase RMS values range from approximately 10-35\,degrees, on average, and the average water vapour measure is approximately 7\,mm. The  {\sc MAPPING} routine was used to produce the calibrated datacubes which could then be output for analysis.  

The large beam sizes of the $^{12}$CO(4--3) observations (Table \ref{tab:beam}), mean that we do not resolve any of the galaxies. Indeed, we expect the $^{12}$CO(4--3) emission regions to be compact, emanating mainly from star forming regions. However, the [C{\sc i}](1--0) emission could be much more extended, tracing the full cold gas distribution as $^{12}$CO(1--0) does. \cite{Ivison11} determine typical $^{12}$CO(1--0) sizes of $\sim$16\,kpc in luminous SMGs. At this size the corresponding  [C{\sc i}](1--0) emission would still remain unresolved in our observations (Table \ref{tab:beam}). The beam shapes are shown in Figs \ref{fig:s7}, \ref{fig:s11}, \ref{fig:s4} and \ref{fig:sa2296}. The  1.4\,GHz positions and redshifts from the new $^{12}$CO(4--3) observations are given in Table \ref{tab:sample} along with the on-source integration times for each line at  their respective observed-frame frequencies.

\begin{table*}
\centering
\begin{tabular}{|l|c|c|c|c|c|c|c|c|c|}
\hline\hline
ID & Beam Size$\rm_{[CI](1-0)}$ & Physical Scale$\rm_{[CI](1-0)}$ & Beam Size$\rm_{CO(4-3)}$ & Physical Scale$\rm_{CO(4-3)}$ \\
\hline
 & (arcsec) & (kpc) & (arcsec) & (kpc) \\
\hline
SXDF7   & 4.4 $\times$ 3.5 & 40 $\times$ 30  & 7.1 $\times$ 3.6 & 60 $\times$ 30 \\
SXDF11  & 8.0 $\times$ 3.2 & 70 $\times$ 30  & 4.4 $\times$ 3.5 & 40 $\times$ 30 \\ 
SXDF4   & 5.1 $\times$ 2.7 & 40 $\times$ 20  & 4.4 $\times$ 3.3 & 40 $\times$ 30 \\
SA22.96 & 4.8 $\times$ 3.2 & 40 $\times$ 30  & 4.5 $\times$ 3.8 & 40 $\times$ 30 \\
\hline\hline
\end{tabular}
\caption{The beam sizes of the [C{\sc i}](1--0) and $^{12}$CO(4--3) observations and the approximate physical scales (in kpc) corresponding to these beam sizes.}
\label{tab:beam}
\end{table*}  

\section{Analysis}
\label{sec:analysis}

\subsection{[C{\sc i}](1--0) and $^{12}$CO(4--3) line extraction}
\label{sec:new}
\begin{figure}
\centering
\includegraphics[width=0.95\textwidth]{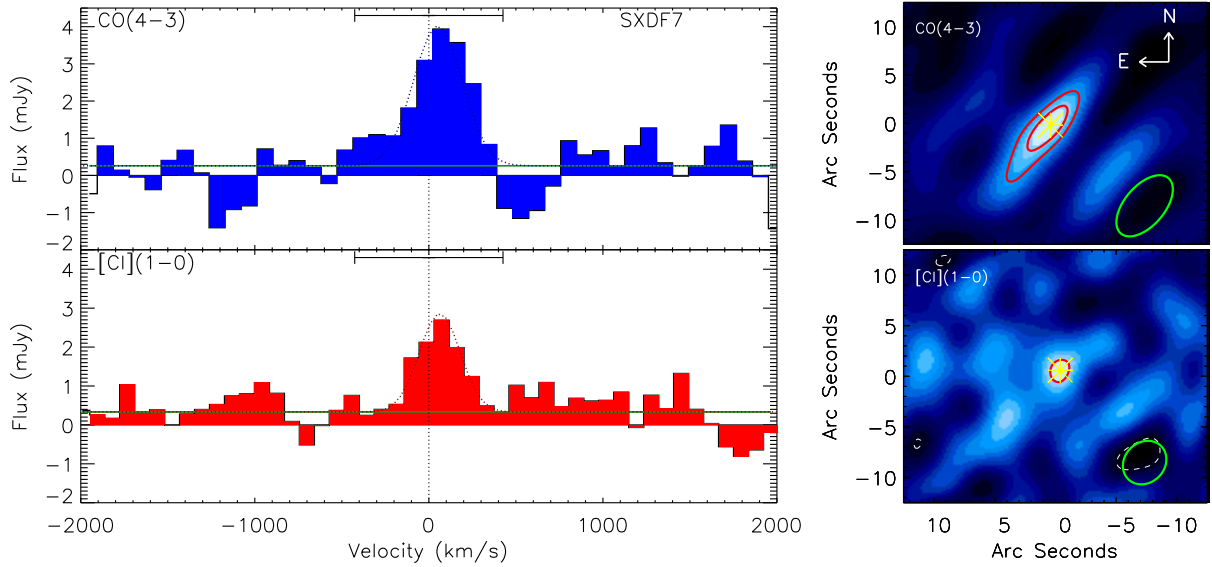}
\caption{SXDF7: Top: $^{12}$CO(4--3) spectrum (left) extracted from the peak pixel of the velocity-integrated S/N map (right) marked by a star.  The contours of the S/N map represent the S/N starting at the $\pm$3$\sigma$ level (solid red for positive $\sigma$ levels and dashed white for negative $\sigma$ levels). The noise is derived from the standard deviation of the flux outside the velocity channels over which the spectra were integrated to measure the intensities. Bottom: [C{\sc i}](1--0) spectrum (left) extracted the pixel marked in the velocity-integrated S/N map (right).  The region marked over the spectra represent the velocity channels over which the spectra were integrated to gain the total line fluxes, as determined from the second order moment analysis of the $^{12}$CO(4--3) spectrum ($\pm2\sigma$). The spectra are centred such that the zero velocity corresponds to the redshift of the $^{12}$CO(4--3) detection. The solid horizontal green lines in the spectra represent the continuum level which was subtracted prior to the moment analysis. The green ellipses on the bottom right of the S/N maps represent the beam sizes and shapes. The maps are centred at  RA=02:17:38.885 and Dec=-05:05:27.971.}
\label{fig:s7}
\end{figure}

\begin{figure}
\centering
\includegraphics[width=0.95\textwidth]{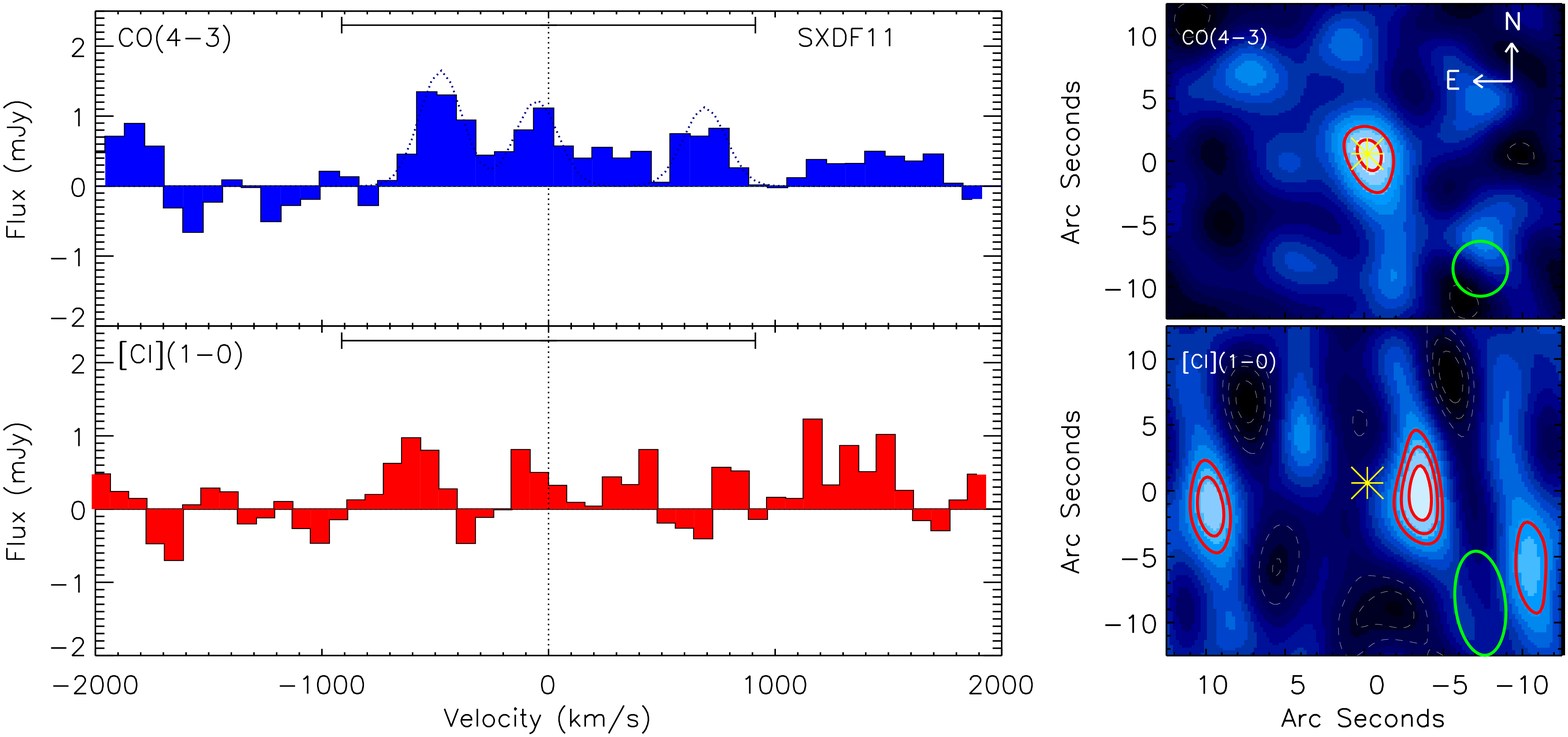}
\caption{SXDF11: Top: $^{12}$CO(4--3) spectrum (left) extracted from the peak pixel of the velocity-integrated S/N map (right) marked by a star.  The contours of the S/N map represent the S/N starting at the $\pm$3$\sigma$ level (solid red for positive $\sigma$ levels and dashed white for negative $\sigma$ levels). The noise is derived from the standard deviation of the flux outside the velocity channels over which the spectra were integrated to measure the intensities. Bottom: [C{\sc i}](1--0) spectrum (left) extracted from the velocity-integrated S/N map (right) at the position of the peak in the $^{12}$CO(4--3) velocity-integrated S/N map, showing that we do not detect [C{\sc i}](1--0) in the same galaxy as the well-centred $^{12}$CO(4--3) detection.  The region marked over the $^{12}$CO(4--3) spectrum represents the velocity channels over which the spectrum was integrated to gain the total line flux, as determined from the second order moment analysis of the $^{12}$CO(4--3) spectrum ($\pm2\sigma$). The spectra are centred such that the zero velocity corresponds to the redshift of the $^{12}$CO(4--3) detection. No continuum was detected therefore we do not subtract continuum from the spectra. The green ellipses on the bottom right of the S/N maps represent the beam sizes and shapes. The maps are centred at  RA=02:17:25.135 and Dec=-04:59:34.296.}
\label{fig:s11}
\end{figure}

\begin{figure}
\centering
\includegraphics[width=0.95\textwidth]{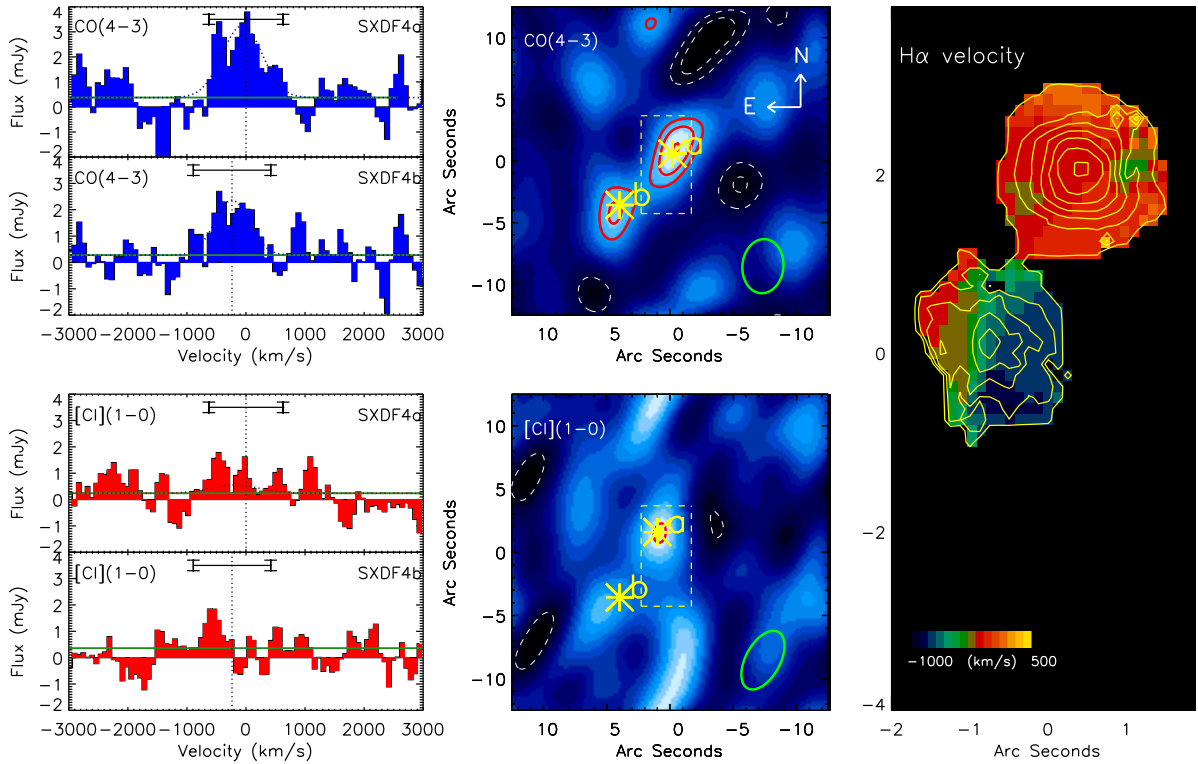} 
\caption{SXDF4: Top: $^{12}$CO(4--3) spectra of the `a' and `b' components of the SXDF4 system (left) extracted from the peak pixels of the velocity integrated S/N map (middle). The contours of the S/N map represent the S/N starting at the $\pm$3$\sigma$ level (solid red for positive $\sigma$ levels and dashed white for negative $\sigma$ levels). The noise is derived from the standard deviation of the flux outside the velocity channels over which the spectra were integrated to measure the intensities. The positions of `a' and `b' are marked on all maps. The regions marked over the spectra represent the velocity channels over which the spectra were integrated to gain the total line fluxes, as determined from the second order moment analysis of the $^{12}$CO(4--3) spectra ($\pm2\sigma$) of `a' and `b'. The spectra are centred such that the zero velocity corresponds to the redshift of the $^{12}$CO(4--3) detection of the `a' component. The S/N maps are created by integrating over the full velocity range of the two components, from the lowest velocity channel of the `b' detection to the highest velocity channel of the `a' detection. Bottom: The [C{\sc i}](1--0) spectrum extracted from the `a' and `b' components, marked by the stars in the velocity-integrated S/N map (middle). The `b' component is not detected in [C{\sc i}](1--0).  The solid horizontal green lines in the spectra represent the continuum level which was subtracted prior to the moment analysis. The green ellipses on the bottom right of the S/N maps represent the beam sizes and shapes. Right: H$\alpha$ velocity field from \protect\cite{Alaghband-Zadeh12} showing distinct multiple components closer in spatial separation than the  $^{12}$CO(4--3) components but further apart in velocity . The yellow dashed rectangle marked on the S/N maps represents the field of view of the H$\alpha$ observations, showing that the `a' component aligns to the $^{12}$CO(4--3) `a' component however the `b' components do not match in position. The maps are centred at  RA=02:17:38.688 and Dec=-05:03:39.29.}
\label{fig:s4}
\end{figure}

\begin{figure}
\centering
\includegraphics[width=0.95\textwidth]{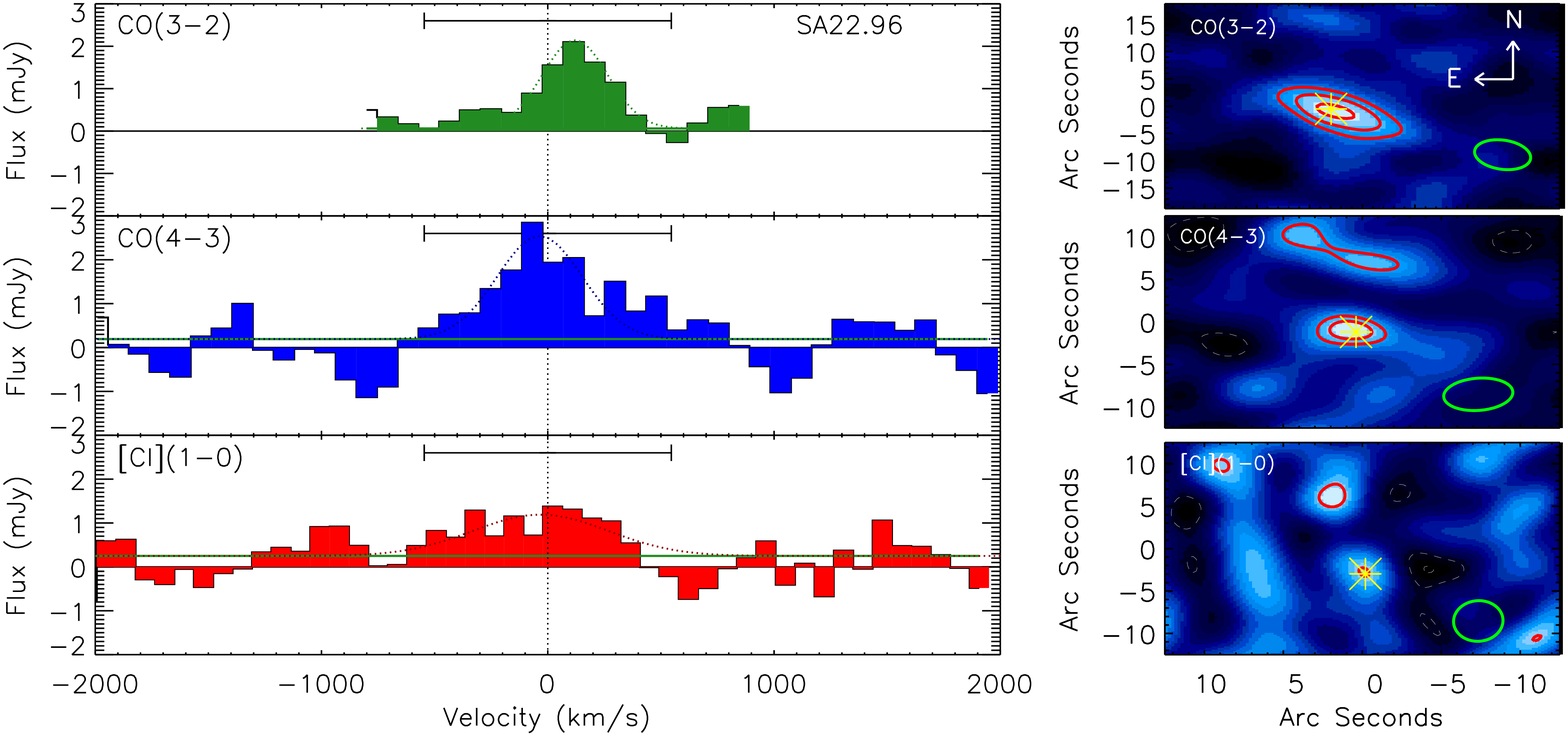}
\caption{SA22.96: Top: $^{12}$CO(3--2) spectrum (left) extracted from pixel marked by a star in the velocity-integrated S/N map (right) from observations detailed in \protect\cite{Bothwell12}. The $^{12}$CO(3--2) observations do not extend to the wide velocity (frequency) range of the $^{12}$CO(4--3) and [C{\sc i}](1--0) observations. The contours of the S/N map represent the S/N starting at the $\pm$3$\sigma$ level (solid red for positive $\sigma$ levels and dashed white for negative $\sigma$ levels). The noise is derived from the standard deviation of the flux outside the velocity channels over which the spectra were integrated to measure the intensities. Middle: $^{12}$CO(4--3) spectrum (left) extracted from the pixel marked in the velocity-integrated S/N map (right). Bottom: [C{\sc i}](1--0) spectrum (left) extracted from the pixel marked in the velocity-integrated S/N map (right). The region marked over the spectra represent the velocity channels over which the spectra were integrated to gain the total line fluxes, as determined from the second order moment analysis of the $^{12}$CO(4--3) spectrum ($\pm2\sigma$). The spectra are centred such that the zero velocity corresponds to the redshift of the $^{12}$CO(4--3) detection. The solid horizontal green lines in the spectra represent the continuum level which was subtracted prior to the moment analysis. The green ellipses on the bottom right of the S/N maps represent the beam sizes and shapes.  The maps are centred at  RA=22:18:4.42 and Dec=00:21:54.40.}
\label{fig:sa2296}
\end{figure}

In Figs. \ref{fig:s7}--\ref{fig:sa2296} we show  signal-to-noise (S/N) maps of the $^{12}$CO(4--3) and [C{\sc i}](1--0) lines. We extract beam averaged, one dimensional spectra from the S/N peaks of the maps. We find that these peaks are clearly associated to faint sources in the IRAC imaging identified by previous radio continuum observations,  as shown in Fig. \ref{fig:cont}.

Continuum was significantly detected in the extra-line regions of the [C{\sc i}](1--0) observations for four of the five SMGs:  SXDF4a, SXDF4b, SXDF7 and SA22.96. In order to measure the continuum at the observed frequency of [C{\sc i}](1--0) in each source, we measure the median flux in the region outside the [C{\sc i}](1--0) lines. This median is shown by the horizontal lines in the extracted one dimensional spectra in Figs. \ref{fig:s7}, \ref{fig:s4} and \ref{fig:sa2296}. We subtract this median level from the spectra before further analysis. We use the measured continuum levels from the [C{\sc i}](1--0) observations to subtract continuum from the detected $^{12}$CO(4--3) sources, since the latter observations (with shorter integration times, and longer wavelengths) do not detect continuum as significantly. The continuum levels are estimated at the rest frequency of $^{12}$CO(4--3) by multiplying the measured continuum at the rest frequency of [C{\sc i}](1--0) by $\rm (\nu_{CO(4-3),rest}/\nu_{[CI](1-0),rest})^{2+\beta}$, where $\beta$=2 \citep{Magnelli12} represents the average spectral energy distribution (SED) of SMGs. Since we do not detect continuum in SXDF11 we do not subtract continuum from either SXDF11 spectrum.

In Figs. \ref{fig:s7}--\ref{fig:sa2296} we show Gaussian fits to the line profiles, often finding multiple components are a better representation of the profile. We show multiple component fits if the fit is improved by allowing more than one Gaussian component. In order to determine the $^{12}$CO(4--3) and [C{\sc i}](1--0) line fluxes and widths we use a second order moment analysis, to accurately represent the full profiles of the lines. First the redshift of the line is calculated as the intensity-weighted centroid of the line, with the line width ($\sigma$) calculated using the intensity-weighted second order moment. We then calculate the flux by integrating between $\pm$2$\sigma$.

The $^{12}$CO(4--3) fluxes, calculated in this way, are given in Table \ref{tab:prop} and the $\pm$2$\sigma$ regions are marked on the spectra in Figs. \ref{fig:s7}--\ref{fig:sa2296}. Since the [C{\sc i}](1--0) detections are weaker than the $^{12}$CO(4--3) detections and the line profiles are not as well defined, we use the $^{12}$CO(4--3) defined velocity range to integrate the [C{\sc i}](1--0) flux. The resulting [C{\sc i}](1--0) fluxes are also given in Table \ref{tab:prop}.

In Table \ref{tab:prop} we bring together the line measurements of our five new SMGs with the literature measurements of [C{\sc i}](1--0) in SMGs from \cite{Walter11}, \cite{Cox11} and \cite{Danielson11}. We also bring together the available $^{12}$CO(1--0), $^{12}$CO(3--2) and $^{12}$CO(4--3) fluxes for this sample. Before discussing the properties of the complete sample we briefly discuss each of the new observations in turn.

\begin{table*}\footnotesize
\centering
\begin{tabular}{|l|c|c|c|c|c|c|c|c|c|c|c|}
\hline\hline
ID & z        & I$\rm_{[CI](1-0)}$ & I$\rm_{[CI](2-1)}$    & I$\rm_{CO(4-3)}$   & I$\rm_{CO(3-2)}$    & I$\rm_{CO(1-0)}$ & References \\
\hline
   &         & (Jy kms$^{-1}$) & (Jy kms$^{-1}$)   & (Jy kms$^{-1}$) & (Jy kms$^{-1}$) & (Jy kms$^{-1}$) & \\
\hline\hline
SXDF7   & 2.529   & 0.8$\pm$0.2    &  -           & 1.6$\pm$0.3    & -              &  -              & 0\\
SXDF11  & 2.282   & $<$0.7         &  -           & 1.0$\pm$0.2    & -              &  -              & 0\\ 
SXDF4a  & 2.030   & 0.7$\pm$0.2    &  -           & 1.1$\pm$0.3    & -              &  -              & 0\\
SXDF4b  & 2.027   & $<$0.7         &  -           & 1.6$\pm$0.5    & -              &  -              & 0\\
SA22.96 & 2.517  & 0.7$\pm$0.2    &  -           & 1.3$\pm$0.3    & 1.6$\pm$0.5    &  -              & 0,3\\
\hline   
J02399  & 2.808   & 1.9$\pm$0.2    & 1.8$\pm$0.9  & -              &  3.1$\pm$0.4   & 0.60$\pm$0.12   &  1,2,9 \\
J123549 & 2.2020  & 1.1$\pm$0.2    & $<$1.0       & -              &  1.68$\pm$0.17 & 0.32$\pm$0.04   &  2,3,4 \\
J163650 & 2.3834  & $<$0.72        & $<$2.0       & -              &  1.97$\pm$0.14 & 0.34$\pm$0.04   &  2,3,4  \\
J163658 & 2.4546  & 0.9$\pm$0.2    & $<$2.0       & -              &  1.5$\pm$0.2   & 0.37$\pm$0.07   &  2,3,4  \\
J14011  & 2.5652  & 1.8$\pm$0.3    & 3.1$\pm$0.3  & -              &  2.8$\pm$0.3   & 0.32$\pm$0.03   &  2,5,11  \\
J16359  & 2.51737 & 1.7$\pm$0.3    & 1.6$\pm$0.3  & 4.0$\pm$0.4    &  2.8$\pm$0.2   &       -         &  2,6  \\
J213511 & 2.3259  & 16.0$\pm$0.5   & 16.2$\pm$0.6 & 17.3$\pm$0.2   &  13.2$\pm$0.1  & 2.16$\pm$0.11   &  7 \\
ID141   & 4.243   & 2.8$\pm$0.9    & 3.4$\pm$1.1  & 7.5$\pm$0.9    &     -          &        -        &  8 \\
MM18423+5938 & 3.930 & 2.3$\pm$0.5 & 4.2$\pm$0.8  & 4.95$\pm$0.05  &   -            & -               &  2,10 \\
\hline\hline
\end{tabular}
\caption{ The redshifts and [C{\sc i}] and $^{12}$CO line fluxes of the complete SMG sample. These fluxes are not corrected for magnification. The SMGs with new [C{\sc i}] detections in this work are in the top of the table. The errors on the fluxes represent the noise in maps in the region outside the source and outside the velocity channel range used to measure the integrated flux. The $^{12}$CO(3--2) flux of SA22.96 is remeasured from the observations of \protect\cite{Bothwell12}. The fluxes in the bottom of the table are the literature fluxes with the following references: 0:This work, 1:\protect\cite{Genzel03}, 2: \protect\cite{Walter11}, 3:\protect\cite{Bothwell12},4:\protect\cite{Ivison11},5:\protect\cite{Downes03},6:\protect\cite{Weiss05},7:\protect\cite{Danielson11}, 8:\protect\cite{Cox11}, 9: \protect\cite{Thomson12}, 10: \protect\cite{Lestrade10}, 11: \protect\cite{Sharon12}. The redshifts quoted for the literature sources come from $^{12}$CO(3--2) where available, or $^{12}$CO(4--3) except for ID141 which is from the combination of both detections and J213511 which comes from the $^{12}$CO(1--0) detection.}
\label{tab:prop}
\end{table*}

\subsubsection{SXDF7}
The S/N contours of SXDF7 shown in Fig. \ref{fig:s7} reveal  clear detections of both $^{12}$CO(4--3) and [C{\sc i}](1--0). Neither line shows any extension beyond the beam shape, and we interpret the source as unresolved in both lines. The offset between the $^{12}$CO(4--3) and [C{\sc i}](1--0) detections of 1.1$''$ is not significant compared to the 7.1$'' \times$3.6$''$  beam size of in the $^{12}$CO(4--3) observations. The H$\alpha$ map of SXDF7 indicates that it comprises two merging components separated by 4$\pm$1\,kpc ($\sim0.5''$) in projection and 100$\pm$50\,kms$^{-1}$ in velocity \citep{Alaghband-Zadeh12}, however owing to the large beam size in these observations compared to the H$\alpha$ observations we cannot determine that the $^{12}$CO(4--3) and [C{\sc i}](1--0) lines are resolved on those scales.

\subsubsection{SXDF11}
The $^{12}$CO(4--3) spectrum of SXDF11 extracted from the peak of the S/N map in Fig \ref{fig:s11} shows multiple velocity components. We use the second order moment analysis to determine the line profile properties. While $5\sigma$, the S/N peak in the [C{\sc i}](1--0) line is offset from the well centred peak (1.0$\pm$0.2 mJy)  of the $^{12}$CO(4--3) line by 3.6$''$, larger than the 3.2$''$ FWHM beam size (east-west) of the [C{\sc i}](1--0) observations, therefore we do not consider this [C{\sc i}](1--0) peak to originate from the same galaxy as the $^{12}$CO(4--3) detection. The $^{12}$CO(4--3) detection is centred on a radio-identified source in the IRAC imaging shown in Fig. \ref{fig:cont}. For further analysis, we treat  [C{\sc i}](1--0) as undetected in SXDF11 and  quote a 3$\sigma$ upper limit to the [C{\sc i}](1--0) flux in Table \ref{tab:prop}. We note, however, that the `offset' source, apparently detected in [C{\sc i}](1--0) and not $^{12}$CO(4--3) may represent a more quiescent star-forming galaxy (and therefore with weak $^{12}$CO(4--3) emission) in a group with the highly star-forming SXDF11 system.

\subsubsection{SXDF4a and SXDF4b}
\label{sec:sxdf4}
The S/N map of $^{12}$CO(4--3) for SXDF4 reveals extended $^{12}$CO(4--3), spanning two $>4\sigma$ components, extending  beyond the beam shape suggesting the source is resolved even with this large beam size (4.4$''$ by 3.3$''$). Two massive components are also seen in the strong H$\alpha$ detections of two high SFR galaxies presented in \cite{Alaghband-Zadeh12} (separated by a velocity of 670$\pm$70\,kms$^{-1}$ and spatially by 18$\pm$5\,kpc). However, the two $^{12}$CO(4--3) components  are  separated by 49$\pm$3\,kpc,  slightly larger than the FHWM of the beam, and in velocity space by 230$\pm$40\,kms$^{-1}$, as shown in Fig. \ref{fig:s4}, significantly larger than the H$\alpha$ spatial separation. This `b' component in $^{12}$CO(4--3) could represent a third source in the complex SXDF4 system; the offset between the southern H$\alpha$ galaxy and the  $^{12}$CO(4--3) `b' position is greater than the CO beam size, and the 'b' component lies just beyond the H$\alpha$ IFU map. 
We consider the two components to be separate SMGs, and extract the spectra from the peak S/N pixel of each component. While there is significant velocity overlap between the beam-averaged broad lines, there is little or no spatial overlap. We cannot rule out that some of the intervening CO emission is coming directly from the second H$\alpha$ galaxy.

We note that the H$\alpha$ maps highlight SXDFa as two highly star-forming galaxies in a major merger \citep{Alaghband-Zadeh12}, providing useful additional diagnostics to the $^{12}$CO maps of the molecular gas, despite H$\alpha$ being heavily extinct in these dusty galaxies \citep{Takata06}. These two H$\alpha$ components are not resolved spatially or in velocity  in our $^{12}$CO(4--3) map, and higher resolution, sensitive $^{12}$CO observations would provide a crucial comparison to the H$\alpha$ map. The combination of H$\alpha$ and $^{12}$CO observations clearly provides a more complete picture of starburst galaxies, as found for instance in studies of local starbursts such as M82 where H$\alpha$ also highlights a significant component of the ionized gas in outflows \citep{Bland88}.

We also detect $\sim$2mm continuum across both components (Table \ref{tab:sample}) and the south-east extension is seen in the IRAC 3.6\,$\mu$m imaging shown in Fig. \ref{fig:cont}. The 2mm continuum in the `a' and `b' components is indistinguishable within the errors, therefore we also split the unresolved 850\,$\mu$m and {\it Herschel}-SPIRE fluxes equally, leading to equal far-infrared luminosities for `a' and `b' .

We note that the [C{\sc i}](1--0) detection in SXDF4a is offset from the $^{12}$CO(4--3) detection by 1.6$''$ however since this is well within the beam size of the [C{\sc i}](1--0) observations we consider the [C{\sc i}](1--0) detection to be associated with the same source as the $^{12}$CO(4--3). [C{\sc i}](1--0) is not significantly detected in SXDF4b therefore we quote a 3$\sigma$ upper limit to the [C{\sc i}](1--0) flux for this component.

\subsubsection{SA22.96}
\begin{figure}
\centering
\includegraphics[width=0.95\textwidth]{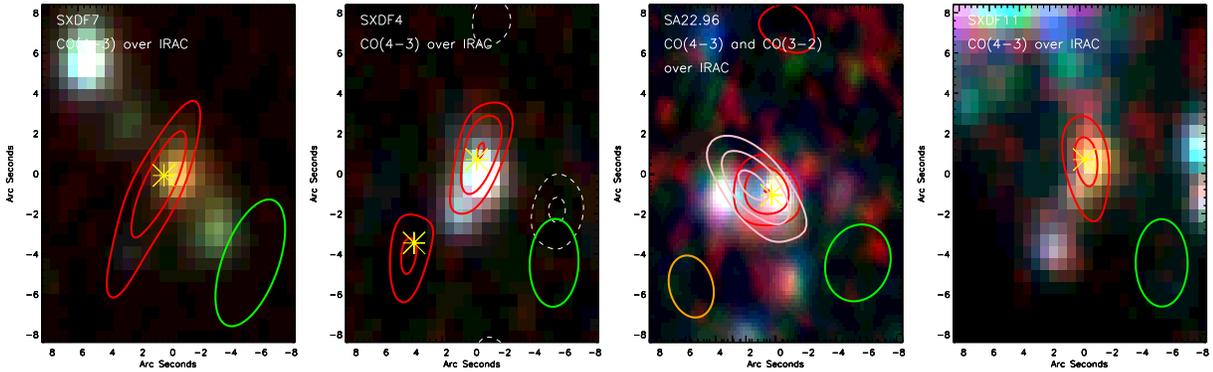}
\caption{Contours of S/N of the $^{12}$CO(4--3) detections (solid red for positive $\sigma$ levels and dashed white for negative $\sigma$ levels), starting at $\pm$3$\sigma$, overlaid on colour IRAC imaging (red, green and blue corresponding to the 3.6\,$\mu$m, 4.5\,$\mu$m and 5.8\,$\mu$m images respectively). The green ellipses on the bottom right represent the beam shapes. We overlay the contours of the $^{12}$CO(3--2) detection of SA22.96 also (in pink) showing the extension to the east, however this eastern component is not detected in continuum, $^{12}$CO(4--3) or [C{\sc i}](1--0). The orange ellipse on the bottom left represents the $^{12}$CO(3--2) beam shape. The stars mark the position of the pixels from which the spectra are extracted. }
\label{fig:cont}
\end{figure}

In Fig. \ref{fig:sa2296} we show the $^{12}$CO(4--3) and [C{\sc i}](1--0) observations of SA22.96 along with a reanalysis of the $^{12}$CO(3--2) observations from \cite{Bothwell12}. The $^{12}$CO(3--2) is clearly extended beyond the beam in this system, suggesting that there may be a second component to the east of the primary source identified by radio continuum and our  $^{12}$CO(4--3) and [C{\sc i}](1--0) detections. A second source in the IRAC imaging  (shown in Fig. \ref{fig:cont}) may be the counterpart to the extended $^{12}$CO(3--2), however this component is not detected in our 2mm continuum map or the 1.4GHz radio map, and we do not separate this potential second component out in the analysis of the $^{12}$CO(3--2) spectrum.

We also note that the centre of the $^{12}$CO(3--2) detection is 2.7$''$ offset to the east from the centre of the $^{12}$CO(4--3) detection, which could again reflect two components contributing to the $^{12}$CO(3--2) flux. However, since the error in the positions of the lines is $\sim$1.3$''$, the two CO transitions are just consistent with still being a single source, and we treat them as such hereafter.

We also note that the H$\alpha$ map centred on the $^{12}$CO(4--3) centroid reveals two  components with $\sim1''$ separation (\citealt{Alaghband-Zadeh12}) demonstrating that SA22.96 is clearly in the process of a major merger, however the H$\alpha$ map does not spatially cover the potential eastern source.

\section{Results}
\label{sec:results}

\subsection{Line Profiles}
\label{sec:shapes}
In this section, we compare  $^{12}$CO and  [C{\sc i}](1--0)  line-widths  within our new sample, and also in the literature sources, to set a context for  whether the lines are tracing the same gas kinematically. In our three sources where [C{\sc i}](1--0) is detected, we compare in Table \ref{tab:sig} the $^{12}$CO(4--3) and  [C{\sc i}](1--0)  line-widths (quantified as $\sigma_{\rm vel}$), now determined independently from a second order moment analysis.

We first apply the second order moment analysis directly on the [C{\sc i}](1--0) detections, which will serve as a verification of our method in Section \ref{sec:new} of determining [C{\sc i}](1--0) fluxes by integrating over velocity channels  determined by fitting the higher S/N $^{12}$CO(4--3) lines. Despite some apparent differences in the velocity centroids from the $^{12}$CO(4--3) lines, we find the [C{\sc i}](1--0) line-widths are still consistent within errors with the  $^{12}$CO(4--3) line-widths in all cases.  The [C{\sc i}](1--0)  fluxes determined directly in this way are also mostly within the errors of the fluxes determined using the $^{12}$CO(4--3) defined velocity channels.  The exception is SXDF4a, where the weak [C{\sc i}](1--0) line surrounded by strong continuum is particularly difficult to measure without reference to $^{12}$CO(4--3). 

The median line-width ratio of $\rm \sigma_{\rm CO(4-3)}$ to $\rm \sigma_{\rm [CI](1-0)}$  is 1.01$\pm$0.09 indicating a good agreement and therefore suggesting that the [C{\sc i}](1--0) may highlight similar gas distributions as the $^{12}$CO(4--3). However, the lower S/N of the [C{\sc i}](1--0) lines implies that we cannot rule out the slightly broader lines (at the 10-20\% level) in this low excitation gas that might be expected based on work by \cite{Ivison11} for the $^{12}$CO(1--0) line in SMGs. 

We then compare the line-widths of four SMGs which have [C{\sc i}](1--0) detections from \cite{Walter11} and $^{12}$CO(1--0) detections from \cite{Ivison11} and \cite{Thomson12}. We apply the second order moment analysis to the line profiles (see Table \ref{tab:sig}) finding that two of the four (J123549 and J163658) have consistent [C{\sc i}](1--0) and $^{12}$CO(1--0) line-widths. However the remaining two (J02399 and J163650) have significantly larger $^{12}$CO(1--0) line-widths compared to our measured [C{\sc i}](1--0) line-widths by $\sim$100\,kms$^{-1}$ in $\sigma$, a $\sim$50\% increase. Our $\rm \sigma_{\rm [CI](1-0)}$  measurements for these two are actually very similar to the higher-J CO lines \citep{Bothwell12}. Indeed, \cite{Ivison11} find that in most cases, the gas probed by $^{12}$CO(1--0) is  more spatially extended than the gas probed by higher-J $^{12}$CO(3--2).

It appears that we cannot conclude that the [C{\sc i}](1--0) line is always tracing the full cold gas component probed by the $^{12}$CO(1--0) line, which may be a result of both optical depth and radiative transfer effects as explored in detail in \cite{Papadopoulos04a}. Further detailed comparisons and resolved studies of $^{12}$CO(1--0) to [C{\sc i}](1--0) are needed to probe this further and understand the origin of the differences. 

However, in order to compare the brightness distributions of these two tracers, and thus the use of each line as H$_2$ mass tracers, the sensitivities of both the [C{\sc i}](1--0) and $^{12}$CO(1--0) moment maps must reach the same gas mass surface density limit. When imaging high-redshift $^{12}$CO(1--0) with the Karl G. Jansky Very Large Array (JVLA) and  [C{\sc i}](1--0) with ALMA, resolution and UV-plane coverage need to be well matched.

\onecolumn

\twocolumn

From this context we next look at the line luminosities and ratios in $^{12}$CO and  [C{\sc i}](1-0).

\begin{table}\footnotesize
\centering
\begin{tabular}{|l|c|c|c|c|c|c|c|c|c|c|c|}
\hline\hline
ID & $\rm \sigma_{\rm [CI](1-0)}$ & $\rm \sigma_{\rm CO(4-3)}$ & $\rm \sigma_{\rm CO(3-2)}$  \\
\hline
   &   kms$^{-1}$ & kms$^{-1}$ & kms$^{-1}$ \\
\hline\hline
SXDF7    & 190$\pm$30  & 220$\pm$60 & - \\
SXDF4a   & 250$\pm$40  &  230$\pm$20 & - \\
SA22.96 & 270$\pm$40  & 270$\pm$30 & 270$\pm$30  \\
\hline
ID & $\rm \sigma_{\rm [CI](1-0)}$  & $\rm \sigma_{\rm CO(1-0)}$ & $\rm \sigma_{\rm CO(1-0)}$ \\
\hline
J02399   & 260$\pm$20 & 380$\pm$10 & - \\
J123549  & 200$\pm$40 & 227$\pm$5 & 230$\pm$20 \\
J163650  & 230$\pm$20 & 320$\pm$10 & 330$\pm$35 \\
J163658  & 260$\pm$20 & 290$\pm$10 & 295$\pm$10 \\
\hline\hline
\end{tabular}
\caption{Comparison the the line-widths ($\sigma$) determined from the second order moment analysis of the $^{12}$CO(4--3) and [C{\sc i}](1--0) detections in our new sample of SMGs (top 3), and in four SMGs (bottom rows) observed in [C{\sc i}](1--0) \protect\citep{Walter11} and $^{12}$CO(1--0) \protect\citep{Ivison11,Thomson12}. The last column compares the line widths measured by \protect\cite{Ivison11} and \protect\cite{Thomson12}.}
\label{tab:sig}
\end{table}

\subsection{Line luminosities and ratios}
\label{sec:profile}
To understand the range of gas properties in SMGs, we bring together the sample of 14 SMGs with [C{\sc i}](1--0) observations, including the five new observations from this work and the nine [C{\sc i}] detections from literature \citep{Walter11,Danielson11,Cox11}, and convert the [C{\sc i}] and $^{12}$CO integrated fluxes to luminosities using the following relation \citep{Solomon05}:

\begin{equation} 
\rm L' (K\,km\,s^{-1} pc^2) = 3.25 \times 10^7 \, I \, \nu_{obs}^{-2}\,  D_L^2 (1+z)^{-3}
\label{eq:flux}
\end{equation}

\noindent where I is the integrated flux in units of Jy\,km\,s$^{-1}$, D$\rm _L$ is the luminosity distance in units of Mpc and $\nu_{\rm obs}$ is the central frequency of the line. 

In the cases where no $^{12}$CO(4--3) detections are available we use the available $^{12}$CO(3--2)  luminosity and a template CO spectral line energy distribution (SLED) from \cite{Bothwell12} to convert the $^{12}$CO(3--2) luminosity to a $^{12}$CO(4--3) luminosity for direct comparison with our new sample.  We also derive $^{12}$CO(1--0) luminosities (from the $^{12}$CO(4--3) luminosities) for the 9 of 14 sources which have not been observed in $^{12}$CO(1--0). The template from \cite{Bothwell12} includes all the recent $^{12}$CO(1--0) detections in SMGs and therefore takes into account that the average SMG  is best represented by a two component medium with a cooler component more widely distributed than a star-forming one, visible only via the high-J CO line emission. For reference, the relevant conversion factors are $\rm r_{32/10}=0.52\pm0.09$ and  $\rm r_{43/10}=0.41\pm0.07$ \citep{Bothwell12}.

The far-infrared luminosities for SXDF7, SXDF11, SXDF4a, SXDF4b and SA22.96 are derived from modified black body SED fits to the available photometry utilising {\it Herschel}-SPIRE fluxes as detailed in \cite{Alaghband-Zadeh12}, and as refined in Section \ref{sec:sxdf4}.  The far-infrared luminosities for the literature sample are taken from \cite{Walter11} and \cite{Cox11} and are corrected by a factor of 1.2$\times$ to represent the luminosity from integrating under a modified blackbody curve from 8--1000\,$\mu$m rather than the 40--400\,$\mu$m quoted in \cite{Walter11}. All luminosities are corrected for gravitational lensing using the magnification factors in Table \ref{tab:lum}. The resulting luminosities are given in Table \ref{tab:lum}.

\begin{table*}\footnotesize
\centering
\begin{tabular}{|l|c|c|c|c|c|c|c|c|c|}
\hline\hline
ID & $ \rm L'_{[CI](1-0)}$ & $ \rm L'_{[CI](2-1)}$&  $ \rm L'_{CO(4-3)}$ &  $ \rm L'_{CO(3-2)}$ &  $ \rm L'_{CO(1-0)}$ & $ \rm L_{FIR}$ & $\mu$\\
\hline
 & (10$^{10}$K kms$^{-1}$ pc$^2$) & (10$^{10}$K kms$^{-1}$ pc$^2$) &(10$^{10}$K kms$^{-1}$ pc$^2$) & (10$^{10}$K kms$^{-1}$ pc$^2$) & (10$^{10}$K kms$^{-1}$ pc$^2$) &(10$^{13}$L$_{\odot}$) & \\
\hline\hline
SXDF7    & 1.3$\pm$0.3 & -                  &  2.9$\pm$0.6   & {\it 3.7$\pm$0.8} &    {\it 7$\pm$2}     & 0.21$\pm$0.05 &     1  \\
SXDF11   & $<$1.0      & -                  &  1.6$\pm$0.3   & {\it 2.0$\pm$0.4} &    {\it 3.9$\pm$0.8} & 0.15$\pm$0.05 &     1  \\
SXDF4a   & 0.7$\pm$0.3 & -                  &  1.4$\pm$0.4   & {\it 1.7$\pm$0.5} &    {\it 3$\pm$1}     & 0.09$\pm$0.02 &     1  \\
SXDF4b   & $<$0.8      & -                  &  2.0$\pm$0.6   & {\it 2.5$\pm$0.8} &    {\it 5$\pm$2} & 0.09$\pm$0.02 &     1  \\
SA22.96 & 1.2$\pm$0.4 & -                  &  2.3$\pm$0.6   & {\it 6$\pm$2}     &    {\it 6$\pm$2}     & 0.5$\pm$0.1   &     1  \\
\hline     
J02399   & 1.5$\pm$0.2   & 0.5$\pm$0.3      &  {\it 3.8$\pm$0.5} & 4.9$\pm$0.6       & 8$\pm$2 {\it (9$\pm$1)}           & 0.7$\pm$0.3   &   2.5 \\   
J123549  & 1.4$\pm$0.3   & $<$0.5           &  {\it 3.4$\pm$0.3} & 4.4$\pm$0.4       & 7.5$\pm$0.9 {\it (8.4$\pm$0.8)}       & 0.5$\pm$0.2   &   1  \\
J163650  & $<$1          & $<$1             &  {\it 4.6$\pm$0.3} & 5.9$\pm$0.4       & 9$\pm$1  {\it (11.3$\pm$0.8)}          & 0.5$\pm$0.2   &   1  \\
J163658  & 1.4$\pm$0.3   & $<$1             &  {\it 3.7$\pm$0.5} & 4.7$\pm$0.6       & 10$\pm$2 {\it (9$\pm$1)}              & 0.7$\pm$0.3   &   1  \\
J14011   & 0.7$\pm$0.    & 0.48$\pm$0.05    & {\it 1.9$\pm$0.2}  & 2.4$\pm$0.3       & 2.4$\pm$0.2{\it 4.5$\pm$0.5}          & 0.2$\pm$0.1   &   4  \\
J16359   & 0.12$\pm$0.02 & 0.043$\pm$0.008  & 0.33$\pm$0.03      & 0.42$\pm$0.03     & {\it 0.80$\pm$0.08}                  & 0.05$\pm$0.02 &   22  \\
J213511  & 0.69$\pm$0.02 & 0.26$\pm$0.01    & 0.85$\pm$0.01      & 1.159$\pm$0.009   & 1.71$\pm$0.09  {\it (2.08$\pm$0.02)} & 0.23$\pm$0.09 &   32.5 \\
ID141    & 0.5$\pm$0.2   & 0.24$\pm$0.08    &  1.6$\pm$0.2       & {\it 2.1$\pm$0.2} & {\it 3.9$\pm$0.5}                   & 0.43$\pm$0.02 &   20  \\
MM18423+5938 & 0.39$\pm$0.08 & 0.26$\pm$0.05 & 0.95$\pm$0.01     & {\it 1.20$\pm$0.01} & {\it 2.31$\pm$0.02}               & 0.23$\pm$0.09 & 20  \\
\hline\hline
\end{tabular}
\caption{Line luminosities and quoted magnification factors for the SMGs studied in this work (top) and the literature SMG sample (bottom). Where certain $^{12}$CO transition line luminosities are not directly measured the values quoted are inferred from other $^{12}$CO transitions using the conversions from \protect\cite{Bothwell12} and are shown in italics.  All luminosities are corrected for gravitational lensing using the magnification factors in column ($\mu$). The far-infrared luminosities are derived from integrating under a modified blackbody curve from 8-1000\,$\mu$m. The magnification value we use for J14011 is 4 since the range of possible magnifications is 3--5 \protect\citep{Smail05}. The magnification value we use for ID141 is 20 since the range of possible magnifications is 10--30 \protect\citep{Cox11}. Since we detect approximately equal continuum across both SXDF4a and SXDFb, the L$\rm _{FIR}$ values for SXDF4a and SXDF4b assume that the observed L$\rm _{FIR}$ is divided equally between the two components of this system. }
\label{tab:lum}
\end{table*}

\begin{figure*}
\centering
\includegraphics[width=0.9\textwidth]{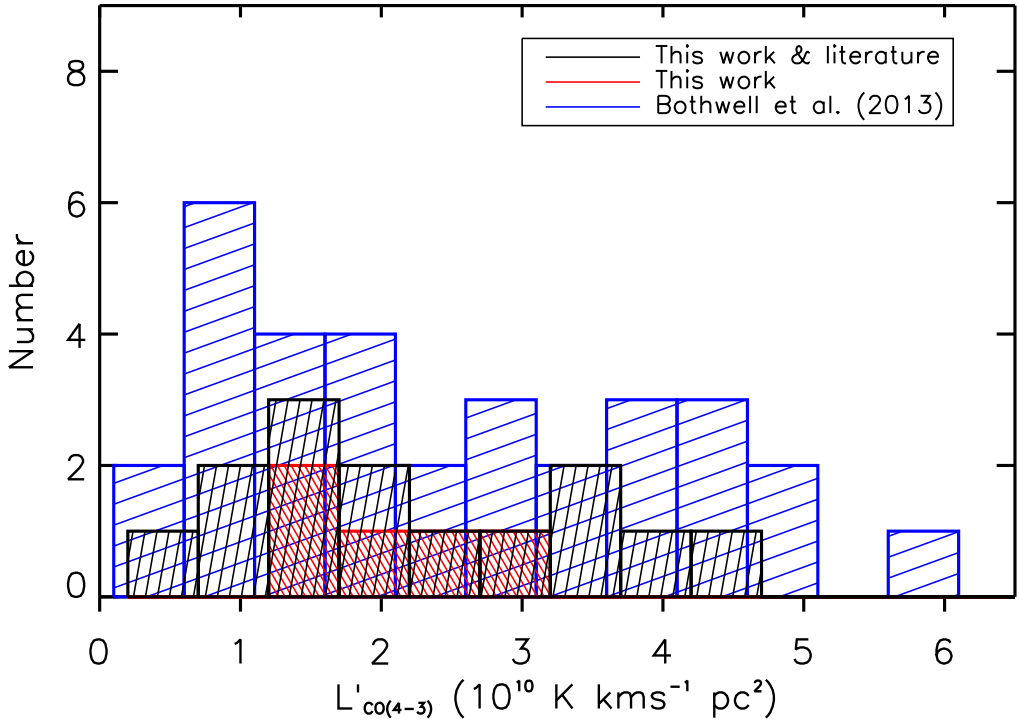}
\caption{The distribution of $\rm L'_{CO(4-3)}$ of the SMGs with new [C{\sc i}] detections in this work (red dense shading) and the combination of this sample and the literature sample of SMGs with [C{\sc i}] detections (black medium shading) compared to the large sample of SMGs which have be studied in $^{12}$CO by \protect\cite{Bothwell12} (blue sparse shading).}
\label{fig:histco43}
\end{figure*}

We first aim to put our SMGs with new [C{\sc i}] detections in this work  in context with the SMGs with literature [C{\sc i}] detections. The two samples have similar median $\rm L'_{CO(4-3)}$:  2.0$\pm$0.4$\times$10$^{10}$\,K kms$^{-1}$pc$^2$  for the SMGs with new detections and 2$\pm$1$\times$10$^{10}$\,K kms$^{-1}$pc$^2$ for the literature SMGs, but the literature sample spans a broader distribution. The combination of both samples probes the $^{12}$CO(4--3) distribution of a large sample of SMGs studied in $^{12}$CO from \cite{Bothwell12} well, as shown in Fig. \ref{fig:histco43}.

\begin{figure*}
\centering
\includegraphics[width=0.8\textwidth]{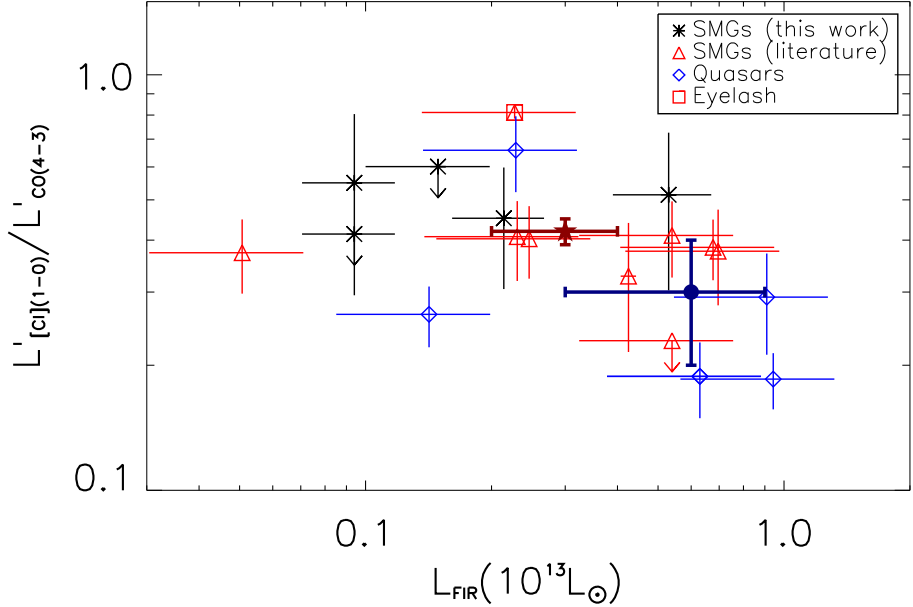}
\caption{Line luminosity ratios and far-infrared luminosities of the SMGs in this work and in the literature compared to the sample of quasars with [C{\sc i}] detections detailed in \protect\cite{Walter11}. All luminosities are corrected for gravitational lensing using the magnification factors in Table \protect\ref{tab:lum}. The ratio of  $\rm L'_{[CI](1-0)}/L'_{CO(4-3)}$ decreases with increasing far-infrared luminosity. The median values for the complete SMG and quasar samples are marked by the dark red filled star and dark blue filled circle respectively. The SMGs extend to higher $\rm L'_{[CI](1-0)}/L'_{CO(4-3)}$ and lower $ \rm L_{FIR}$ values than the quasar sample.  The ratio of $\rm L'_{[CI](1-0)}/L'_{CO(4-3)}$ decreases with increasing $ \rm L_{FIR}$. Assuming that the [C{\sc i}] provides a tracer of the bulk H$_2$ gas mass, in merger systems (with higher $ \rm L_{FIR}$ values) a larger fraction of bulk H$_2$ gas will be participate in the star-formation therefore the ratio of $\rm L'_{[CI](1-0)}/L'_{CO(4-3)}$ will be reduced compared to systems with extended distributions of star-formation idle H$_2$ gas which will have higher values of  $\rm L'_{[CI](1-0)}/L'_{CO(4-3)}$ and lower $ \rm L_{FIR}$ values.}
\label{fig:prop}
\end{figure*}

\begin{figure*}
\centering
\includegraphics[width=0.8\textwidth]{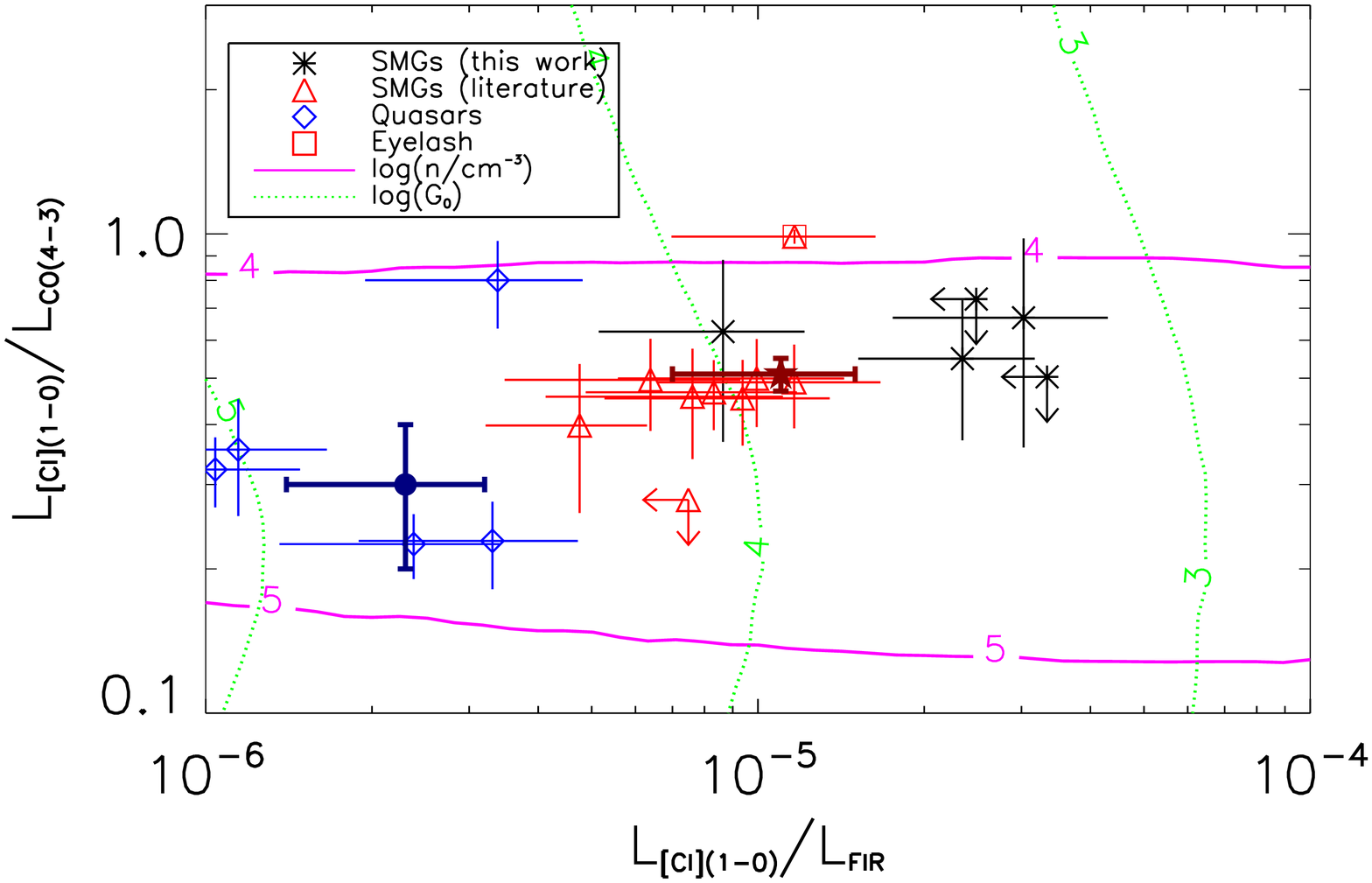}
\caption{Luminosity ratios  (in units of $ \rm L_{\odot}$) of the SMGs in this work and in the literature compared to the quasar sample  with [C{\sc i}] detections detailed in \protect\cite{Walter11}. The median values for the complete SMG and quasar samples are marked by the dark red filled star and dark blue filled circle respectively. The SMGs have higher $ \rm L_{[CI](1-0)}/L_{FIR}$ and $ \rm L_{[CI](1-0)}/L_{CO(4-3)}$ ratios than the literature sample of quasars. The contours plotted indicate the derived values of the gas density ($n$) and the radiation field ($G_0$) for the corresponding ratios of $ \rm L_{[CI](1-0)}/L_{CO(4-3)}$ and $ \rm L_{[CI](1-0)}/L_{FIR}$ from PDR model outputs as detailed in Section \protect\ref{sec:pdr}.  However, we note that if only a small fraction of the [C{\sc i}] emission is due to the PDRs (Section \ref{sec:pdrissues}) then all the ratio values should be lower than plotted (since the non-PDR contribution to the [C{\sc i}] emission should be subtracted from the observed values) therefore suggesting higher densities and radiation field values for the PDR-related gas. Furthermore, assuming that the [C{\sc i}] is non-PDR distributed and instead traces the  bulk H$_2$ gas mass this plot shows compact starbursts in the low  $ \rm L_{[CI](1-0)}/L_{FIR}$ and $ \rm L_{[CI](1-0)}/L_{CO(4-3)}$ region since the $^{12}$CO(4--3) and far-infrared luminosities trace the same gas as the [C{\sc i}], and systems with extended gas distributions in the high $ \rm L_{[CI](1-0)}/L_{FIR}$ and $ \rm L_{[CI](1-0)}/L_{CO(4-3)}$ region where the [C{\sc i}] traces the star-formation idle H$_2$ gas but the $^{12}$CO(4--3) and far-infrared trace the star-forming gas only. }
\label{fig:prop3}
\end{figure*}

We also compare the median values of line luminosity ratios of the two sub-samples. The median $\rm L'_{[CI](1-0)}/L'_{CO(4-3)}$ ratio is 0.51$\pm$0.05 for the SMGs  with new detections and 0.47$\pm$0.03 for the literature SMGs,  again indicating consistency, and as shown in Fig. \ref{fig:prop}.  There is, however, a marked difference in $\rm L_{[CI](1-0)}/L_{FIR}$ between the two samples; the SMGs with new [C{\sc i}] detections have a median ratio of $2.6\pm0.5\times10^{-5}$ whereas the literature sample has a median ratio of $0.8\pm0.1\times10^{-5}$, as shown in Fig. \ref{fig:prop3}, which provides initial evidence that the properties of the ISM in the SMGs varies, possibly suggesting that some of the SMGs with new detections have larger cool gas reservoirs for their far-infrared luminosity. This difference  may be reduced with better constrained [C{\sc i}] measurements for those SMGs which currently only have upper limits. We combine these two sub-samples into one  sample of 14 SMGs for the remainder of the paper.

We next compare the luminosity ratios of the SMGs to a sample of other high-redshift systems, comprised of bright quasars and radio-loud galaxies  which have [C{\sc i}] detections in \cite{Walter11}.  We do not include BRI1335-0417 from \cite{Walter11} in the quasar sample since [C{\sc i}](1--0) is undetected and we have no direct observations of the low-J $^{12}$CO transitions. We measure a larger median $\rm L_{[CI](1-0)}/L_{CO(4-3)}$ in the SMGs of 0.51$\pm$0.04 compared to 0.3$\pm$0.1 in the quasar sample which can be seen in Fig. \ref{fig:prop3}. We measure a median $\rm L_{[CI](1-0)}/L_{FIR}$ of $1.1\pm0.4\times10^{-5}$ and $0.23\pm0.09\times10^{-5}$ in the SMG sample and the quasar sample respectively which indicates a difference (of $\sim$2$\sigma$) between the two samples in terms of $\rm L_{[CI](1-0)}/L_{FIR}$ which is owing to the larger far-infrared luminosities measured in the quasar systems acting to reduce the value of this ratio. Applying a Kolmogorov-Smirnov test, we derive a 96 per cent probability that the SMGs and quasars were drawn from different populations in terms of $\rm L'_{[CI](1-0)}/L'_{CO(3-2)}$ and therefore we disagree with the findings of \cite{Walter11} since they find no difference between the SMGs and QSOs in terms of $\rm L'_{[CI](1-0)}/L'_{CO(3-2)}$.

Finally, we compare the luminosity ratios to local galaxies to test if the [C{\sc i}] properties differ at lower redshift  in a variety of different ISM environments, using the above conversions to $\rm L'_{CO(1-0)}$ where necessary. We find a median $ \rm L'_{[CI](1-0)}/L'_{CO(1-0)}$ of 0.20$\pm$0.08 for the SMG sample which is consistent with the value derived in a sample of irregular, interacting and merging, nearby galaxies (\citealt{Gerin00}: $ \rm \langle L'_{[CI](1-0)}/L'_{CO(1-0)}\rangle $=0.2$\pm$0.2) and also consistent with the ratio measured in the Milky Way (\citealt{Fixsen99}: $ \rm \langle L'_{[CI](1-0)}/L'_{CO(1-0)}\rangle $=0.15$\pm$0.1). We therefore find the ratio of [C{\sc i}] to $^{12}$CO(1--0) luminosity is similar in galaxies at low and high redshift,  in agreement with \cite{Walter11}, which may lend further support to the idea that [C{\sc i}]  is not completely constrained to the thin interface regions of PDRs.

We note that Fig. \ref{fig:prop} shows a decrease in $\rm L'_{[CI](1-0)}/L'_{CO(4-3)}$ with increasing far-infrared luminosity. This relationship can be explained assuming that the [C{\sc i}] provides a tracer of the bulk H$_2$ gas mass and that $\rm L_{FIR}$  traces the star-forming ISM, since in merger systems with higher $\rm L_{FIR}$, a larger fraction of H$_2$ will be participating in the star-formation (traced by $^{12}$CO(4--3)) such that the ratio $\rm L'_{[CI](1-0)}/L'_{CO(4-3)}$ will be reduced. By comparison, systems with extended distributions of cooler reservoirs of H$_2$ gas will have higher values of  $\rm L'_{[CI](1-0)}/L'_{CO(4-3)}$ and lower $\rm L_{FIR}$. Furthermore, assuming that the [C{\sc i}] provides a tracer of the bulk H$_2$ gas mass, Fig. \ref{fig:prop3} can be interpreted as revealing compact starbursts in the low  $ \rm L_{[CI](1-0)}/L_{FIR}$ and $ \rm L_{[CI](1-0)}/L_{CO(4-3)}$ region since the $^{12}$CO(4--3) and $\rm L_{FIR}$ trace the same gas as the [C{\sc i}]. The higher $ \rm L_{[CI](1-0)}/L_{FIR}$ and $ \rm L_{[CI](1-0)}/L_{CO(4-3)}$ region of the plot then represents systems with extended, cool gas distributions traced by the [C{\sc i}] line, but the $^{12}$CO(4--3) and $\rm L_{FIR}$ trace mainly the actively star-forming gas.

\subsection{PDR models}
\label{sec:pdr}
In order to explore the properties of the ISM within the SMGs we compare the line luminosity ratios to the outputs of molecular cloud models. The ISM of galaxies contains bound molecular clouds and cold neutral gas. Photodissociation regions (PDRs) are regions of the neutral gas where far-ultraviolet photons dominate the structure, heating and chemistry of the ISM \citep{Hollenbach99}. PDR models are an important tool to understand the effects of the far-ultraviolet radiation in all the regions containing atomic and molecular gas \citep{Hollenbach99}. The PDR models of  \cite{Kaufman99} assume a slab geometry for molecular clouds being illuminated from one side and probe the atomic gas density ($n$) and the far-ultraviolet radiation field as a function of the various line luminosity ratios.  We note that there are a number of issues with using these simple one-dimensional PDR models, which we discuss further in Section \ref{sec:pdrissues}, however deriving an estimate of the characteristic density and far-ultraviolet radiation field for the SMGs and the quasar sample provides a useful comparison between the samples even if the output derived values are not true representations of the densities and radiation fields. These average properties can be strongly affected by the different conditions of starburst versus quiescent environments therefore they provide a useful insight into the differences in these galaxies. Furthermore as more line observations are used to constrain these averages the prevailing conditions are further refined \citep{Rangwala11}.

We use our line luminosity ratios together with the far-infrared luminosities to constrain properties of the systems by comparing the ratios to the values output from the \cite{Kaufman99} models. We quote the UV radiation field in multiples of $G_0$, the UV radiation field in the Milky Way in Habing units (1.6$\times$10$^{-3}$\,erg s$^{-1}$ cm$^{-2}$). The models assume a given set of model parameters, including elemental abundances and also dust properties and then solve for the chemistry, thermal balance and radiative transfer in the PDR. A given PDR model has a constant density and radiation field therefore at each density and radiation field the various emitted line-luminosity ratios can be established.

In Figs. \ref{fig:model} and \ref{fig:model_lit} we show the PDR model output for our measured line ratios at different values of $n$ and $G_0$ taken from the models of  \cite{Kaufman99}\footnote{\url{dustem.astro.umd.edu/pdrt/index.html}}. We created finer grids than provided by \cite{Kaufman99},  interpolating between the steps in $n$ and $G_0$. Since the contours of $ \rm L_{[CI](1-0)}/L_{CO(4-3)}$  and  $ \rm L_{[CI](1-0)}/L_{FIR}$ are orthogonal it is possible to define an unique value of $n$ and $G_0$ for each SMG by finding the intersection of the contours. The derived values are shown in Table \ref{tab:ng}.

\begin{table}
\centering
\begin{tabular}{|l|c|c|c|c|c|c|c|c|c|}
\hline\hline
ID & log($n$ (cm$^{-3}$)) & log($G_0$) \\
\hline\hline
SXDF7   & 4.3$\pm$0.1   &  3.5$\pm$0.1 \\
SXDF11  & $$4.1         &  $>$3.4 \\ 
SXDF4a  & 4.2$\pm$0.2   &  3.4$\pm$0.1 \\
SXDF4b  &  $>$4.3       &  $>$3.3 \\
SA22.96 & 4.2$\pm$0.1   &  4.0$\pm$0.1 \\
J02399   &  4.37$\pm$0.09   &   4.1$\pm$0.2  \\
J123549  &  4.4$\pm$0.1   &   4.1$\pm$0.2 \\
J163650  &  $>$4.7          &   $>$4.1 \\
J163658  &  4.4$\pm$0.1     &   4.2$\pm$0.2 \\
J14011   &  4.4$\pm$0.1   &     4.0$\pm$0.2  \\
J16359   &  4.4$\pm$0.1   &     4.1$\pm$0.2  \\
J213511  &  4.1$\pm$0.3   &   3.6$\pm$0.7 \\
ID141    &  4.5$\pm$0.2     &     4.4$\pm$0.2  \\ 
MM18423+5938 & 4.4$\pm$0.1   &     4.3$\pm$0.2  \\
\hline\hline
\end{tabular}
\caption{The inferred values of the density ($n$) and radiation field ($G_0$) for the SMGs. The values quoted for J213511 (the 'Eyelash') are from the analysis of multiple line ratios from \protect\cite{Danielson11}. We use only two line ratios to constrain the densities and radiation field values noting that using more line ratios will increase the spread of the possible solutions such that the typical model errors on log($n$ (cm$^{-3}$)) and log($G_0$) are  0.3 and 0.7 respectively \protect\citep{Danielson11}. }
\label{tab:ng}
\end{table}

\begin{figure*}
\centering
\includegraphics[width=0.95\textwidth]{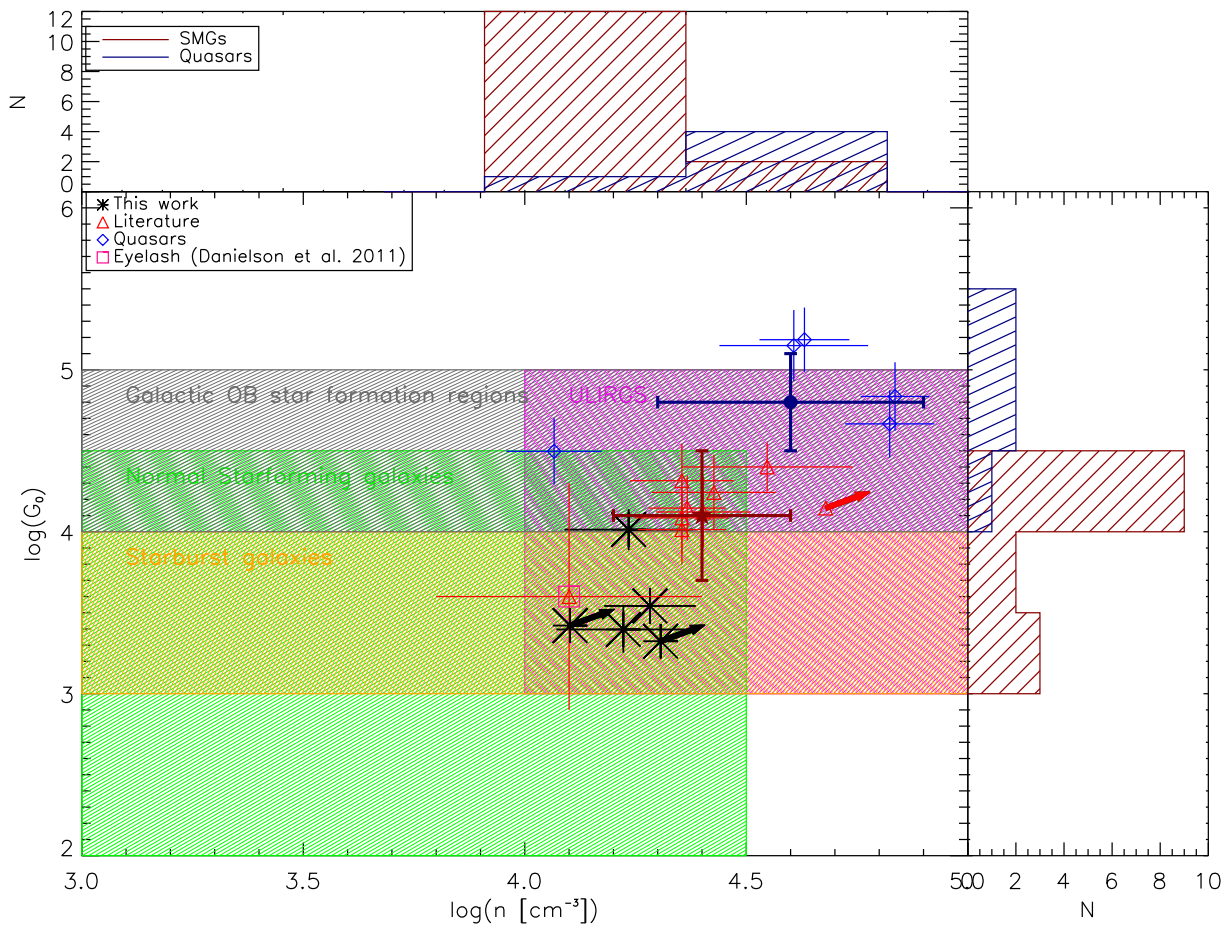}
\caption{The far-UV radiation field, in terms of the local value ($G_0$), against the gas density ($n$) derived from the comparison of various [C{\sc i}], $^{12}$CO and far-infrared luminosities to the outputs from the \protect\cite{Kaufman99} PDR models. We plot the derived values for the SMGs and the literature quasars. The median values for the complete SMG and quasar samples are marked by the dark red filled star and dark blue filled circle respectively. We also mark the ranges in $n$ and $G_0$ derived for other populations - local ULIRGs, Galactic OB star-formation regions, normal star-forming galaxies and starburst galaxies (\protect\citealt{Stacey91,Malhotra01,Davies03}). We find that the SMGs are most comparable to the local ULIRGs although the tail of 5 SMGs with lower densities and radiation field values are more consistent with local starburst galaxies.}
\label{fig:ngplot}
\end{figure*}

In order to compare the $n$ and $G_0$ values of the SMGs to other high-redshift systems and also local systems we plot the distribution of $n$ and $G_0$ in Fig. \ref{fig:ngplot}. We first note that the radiation fields of the literature sample of SMGs are higher than the SMGs with new detections in this work. We measure a median density and radiation field of  $\log (n/\rm cm^{-3})=4.4\pm0.2$ and $\log(G_0)=4.1\pm0.2$ respectively in the literature SMG sample compared to $\log (n/\rm cm^{-3})=4.23\pm0.08$ and $\log(G_0)=3.4\pm0.3$ for the SMGs with new detections in this work. Six of the nine literature SMGs are gravitationally lensed  and it is possible that the densities and radiation fields may be over-estimated owing to differential magnification in these lensed systems. For instance, this may manifest if [C{\sc i}](1--0) probed a larger region of cold gas than the more centrally concentrated region probed by the far-infrared and $^{12}$CO(4--3) observations. Stronger, differential lensing of the far-infrared and $^{12}$CO(4--3) luminosities over [C{\sc i}] would result in higher derived density and radiation field values.

Our line profile analysis (Section \ref{sec:shapes}) does not provide direct evidence for such a scenario for  [C{\sc i}], but we do not spatially resolve the lines in this study, nor is our S/N sufficient to rule out line width differences in the $^{12}$CO and [C{\sc i}] lines.  This scenario could clearly happen for the spatially extended $^{12}$CO(1--0) reservoirs found in \cite{Ivison11}.  We also note that spatially resolved observations reaching the same H$_2$ gas mass surface density sensitivities of both $^{12}$CO(1--0) and  [C{\sc i}](1--0) are required to explore this further. We assume here that differential lensing does not have a large effect on our results, and combine all 14 SMGs into a single sample, finding a median density of $\log (n/\rm cm^{-3})=4.4\pm0.2$ and a median radiation field of  $\log(G_0)=4.1\pm0.4$. We note that the quoted dispersions of the density and radiation fields characterise the distribution of the population, but do not include the additional broadening from measurement errors. Including the average measured errors in the estimate of the SMGs' distributions would reduce the error on log($n$ (cm$^{-3}$)) to 0.07 and log($G_0$) to 0.3. We also note that using more than two line ratios will also increase the spread of the possible solutions as shown in \cite{Danielson11} for the `Eyelash' (J213511) for which numerous line ratios are used (from the $^{12}$CO ladder, [C{\sc i}], [C{\sc ii}] ) giving model errors on log($n$ (cm$^{-3}$)) and log($G_0$) of $\sim$0.3 and $\sim$0.7 respectively

In order to explore the differences in ISM conditions in a range of high-redshift systems we also constrain $n$ and $G_0$ for the quasar sample from \cite{Walter11} as shown in Fig. \ref{fig:model_lit_nonsmg}, finding a median density of $\log (n/\rm cm^{-3})=4.6\pm0.3$ and radiation field of  $\log(G_0)=4.8\pm0.3$.  We therefore derive higher radiation fields in the quasar sample than in the SMGs which may provide evidence that the quasar systems contain more compact starbursts (the ratios of $ \rm L_{[CI](1-0)}/L_{FIR}$ and $ \rm L_{[CI](1-0)}/L_{CO(4-3)}$ are lower, compared to the SMGs), as shown in Fig. \ref{fig:prop3} and Section \ref{sec:profile}. The locus of the quasar properties appears to be well within the local ULIRG regime, and the outlier, as shown in Fig. \ref{fig:ngplot}, may be representative of the star forming SMG host dominating over the AGN in this system, providing support for the interpretation of quasars as transition objects. The gas densities we measure in SMGs are also lower than found for a $z$=6.42 quasar, $n=10^5 \rm cm ^{-3}$ \citep{Maiolino05}, providing additional support for varied gas properties across the high-redshift systems.

In order to assess if the ISM properties in star forming galaxies have varied since the present day, we compare the SMGs to local systems in Fig. \ref{fig:ngplot}. \cite{Davies03} studied local ULIRGs, finding density values between $n=10^4 - 10^5$\,cm$^{-3}$ and radiation fields of $G_0>10^3$ which overlaps with the values we derive for distant SMGs.  However, we note that there is recent evidence that in some local ULIRGs the far-UV photons (and therefore the PDRs) are unable to power the ISM energetics, and therefore the high-J $^{12}$CO lines. The high densities and metal-rich environments in these systems stop the far-UV photons from travelling far, and only cosmic rays and/or turbulent heating can explain the large amounts of dense and warm gas observed in the local ULIRGs \citep{Bradford03,Papadopoulos12b}. Recent {\it Herschel} observations have also shown that local ULIRGs can be powered by shocks and/or X-rays. \cite{VanderWerf10} find that X-ray heating can power the high-J $^{12}$CO lines in Mrk231  and \cite{Hailey-Dunsheath12} favour a X-ray heating model for NGC 1068 and conclude that far-UV heating is unlikely, whereas \cite{Meijerink13} find that shocks are dominating the $^{12}$CO excitation in NGC 6240. 
 
Furthermore, a sample of nearby, normal star forming galaxies is found to have $n$ and $G_0$ values in the ranges $n=10^2 - 10^{4.5}$\,cm$^{-3}$ and $G_0=10^2 - 10^{4.5}$ \citep{Malhotra01} which overlap in the far upper ranges with the values we find for SMGs. The mean density for the SMGs is also consistent with the range derived in Galactic OB regions however the mean radiation field for the SMGs is at the lower range of the  values derived in Galactic OB regions \citep{Stacey91}. Finally the density and radiation field values for the SMGs are consistent with nearby starburst galaxies \citep{Stacey91}, although this belies the fact that half of the 14 SMG sample actually lie significantly above this range in $G_0$.

Overall, we find the  SMGs are most comparable to local ULIRGs in $G_0$ and $n$, however a significant tail of 5 of the 14 SMGs are likely best compared to local starburst galaxies which exhibit lower densities and weaker radiation fields.

\subsubsection{PDR model Issues}
\label{sec:pdrissues}
There are a number of issues associated with PDR modelling since a galaxy is much more complicated than a single PDR characterized by one set of parameters. Firstly, the PDR model uses micro turbulence (small velocity gradients) to estimate the line opacities. These PDR models often fail to model the $^{12}$CO ladder and [C{\sc i}] lines owing to the problems with the small velocity gradients not being altered in the PDR grids. Changing the velocity gradients, or indeed the assumed density profiles, in the PDR models will therefore act to alter the density and radiation field values derived.  We note that not all PDR models use micro turbulence with  some using a large velocity gradient escape probability formalism such as the work by \cite{Jansen95}.

The one dimensional PDR slab models produce a relatively thin layer of neutral atomic gas in between the transition from the C+ rich outer layer to the CO rich inner volume in the FUV-illuminated molecular cloud. This would imply that [C{\sc i}] cannot trace the bulk H$_2$ mass which contrasts with extensive observational evidence that suggests that [C{\sc i}] is fully concomitant with the CO-bright H$_2$ \citep{Plume95,Keene96,Ojha01,Ikeda02}. Indeed \cite{Keene96} note that the simple plane-parallel PDR models of, for example, \cite{Tielens85} do not predict the observed proportionality between [C{\sc i}] and other H$_2$ tracers, or the extended distributions of [C{\sc i}]. This is further illustrated in the attempts to model the well-studied Barnard 68 dark globule by \cite{Pineda07} where the relative [C{\sc i}], $^{12}$CO and  $^{13}$CO cannot be reproduced simultaneously by the spherically symmetric \cite{Storzer96} PDR model with differences between the observations and models of up to a factor of 2. 

There has been a considerable effort to alter the simple models to explain these observations by using low-density PDRs and inhomogeneous PDRs \citep{Hollenbach91,Meixner93,Spaans96}. \cite{Stutzki88} suggest that the surface layers of [C{\sc i}] are distributed across the clouds since the clouds are clumpy. \cite{Spaans96} presents a homogeneous model and a clumpy model with the same (average) density. The average abundances as a function of extinction are identified for both the models, finding that the homogeneous model indeed shows the layered structure, while the clumpy model shows approximately a constant average abundance for C+ and C. The C abundance is found to follow the average H$_2$ abundance well, indicating that the global H$_2$ can be traced with [C{\sc i}]. The \cite{Spaans96} PDR models is found to be successful at predicting the spatial correlation between [C{\sc i}] and other H$_2$ mass tracers such as  $^{12}$CO and $^{13}$CO.  However, the tight relation observed between [C{\sc i}] and $^{13}$CO over a wide range of ISM conditions remains a problem unless the [C{\sc i}] distribution is mostly concomitant  with the bulk CO-rich H$_2$ gas (rather than being locked into the PDRs). The clumpy models allow the far-UV to penetrate deeper into the molecular clouds, however in these models the [C{\sc i}] distribution remains inconsistent with the widespread [C{\sc i}] emission observed deep in molecular clouds with little far-UV illumination \citep{Keene85,Tatematsu99,Oka01}. The clumpy models also cannot achieve the high $^{12}$CO/$^{13}$CO ratios observed in ULIRGs. Furthermore, \cite{Keene85} present observations of [C{\sc i}] in two Galactic molecular clouds finding that the [C{\sc i}] is widely distributed in the clouds and not consistent with only appearing at the cloud edges, close to the ionisation fronts. \cite{Papadopoulos04a} further explore the non-PDR distribution of [C{\sc i}], highlighting observations such as  [C{\sc i}] emission from UV-protected regions in the Galaxy and the similarities between the [C{\sc i}] and  $^{13}$CO line profiles, from which they conclude that observations are best explained by [C{\sc i}] tracing the H$_2$ gas mass rather than having a PDR-like distribution.

We note that there is significant scatter between different  PDR models currently used, as discussed in \cite{Rollig07}. The models contain numerous uncertainties given the various unknown chemical processes which are still being determined and understood experimentally and theoretically. There is therefore an on-going effort to improve the models and converge on a common solution, and further observations are needed to help to guide this.

We conclude that in reality the ISM is likely a clumpy structure with lower and higher density regions which are all caught in the same beam in our observations and given the information available to us, namely the [C{\sc i}](1--0),  $^{12}$CO(4--3) and far-infrared luminosities, we are unable to use more complicated models since that would introduce more parameters than we can constrain with only these line luminosities. Nevertheless, our estimation of the densities and radiation fields of the SMG and quasar samples, from comparison to the simple PDR models, provides us with a useful initial comparison and drives further studies into modelling the molecular clouds.

\subsection{Gas masses}
\label{sec:mass}
There is evidence that the mass derived from the [C{\sc i}] line luminosity can be used to estimate the H$_2$ mass in local ULIRGS \citep{Papadopoulos04b}, Arp220 \citep{Gerin98} and the Cloverleaf QSO \citep{Weiss03}. \cite{Papadopoulos04a} note that the small optical depths of the [C{\sc i}] lines make them ideal tracers of H$_2$ gas, along with the simple partition function structure  and the apparent concomitance of [C{\sc i}] with the bulk H$_2$ gas mass. In order to test this in high-redshift systems we first calculate the total mass of neutral carbon in the SMGs using:

\begin{equation} 
\rm M_{[CI]}= 5.706 \times 10^{-4} Q(T_{ex})\frac{1}{3}e^{23.6/T_{ex}} L'_{[CI](1-0)}[M_{\odot}]
\label{eq:massW}
\end{equation}

\noindent \citep{Weiss05}, which assumes the [C{\sc i}] line is optically thin and where $\rm Q(T_{ex})=1+3e^{-T_1/T_{ex}}+5e^{-T_2/T_{ex}}$ (the [C{\sc i}] partition function). $\rm T_1$ and $\rm T_2$ are the energies above the ground state, 23.6\,K and 62.5\,K respectively and $\rm T_{ex}$ is the excitation temperature. Since we do not have information on the [C{\sc i}](2--1) line for the majority of the sources we cannot constrain the excitation temperature \citep{Schneider03} therefore we take $\rm T_{ex}=$30\,K, consistent with the average derived $\rm T_{ex}$ from \cite{Walter11} mainly from quasars with both [C{\sc i}] line detections ($\rm T_{ex}=29.1\pm6.3$\,K), under the assumption that both lines are optically thin and have the same excitation temperature.  For densities as high as $\sim$10$^4$\,cm$^{-3}$ [C{\sc i}] will be thermalised and therefore local thermodynamical equilibrium (LTE) can be assumed, justifying using the same  $\rm T_{ex}$ for all levels of [C{\sc i}], such that $\rm T_{ex}$ = $\rm T_{kin}$. We derive a median neutral carbon mass of 1.2$\pm$0.2$\times \rm 10^7M_{\odot}$ in the SMGs (Table \ref{tab:mass}). 

We then compare these masses to the molecular gas masses derived from the $^{12}$CO luminosities. We use both the measured $^{12}$CO(1--0) fluxes (where available -- Table \ref{tab:lum}) and the derived $^{12}$CO(1--0) luminosities and assume a conversion of $\rm \alpha_{CO}=1.0\,M_{\odot} (K km s^{-1} pc^2)^{-1}$ such that $\rm M_{CO}(H_2)=\alpha_{CO} L'_{CO(1-0)} [M_{\odot}]$ as used by \cite{Bothwell12},  the use of which we discuss further in Section \ref{sec:phi}. We then measure a median molecular gas mass (from $^{12}$CO(1--0)) of $50\pm10\times 10^9\rm M_{\odot}$  in the SMGs. 

With knowledge of these two masses we can estimate the neutral carbon abundance relative to H$_2$ \citep{Weiss05} using:

\begin{equation} 
\rm X[CI]/X[H_2]=M[CI]/(6 M[H_2])
\label{eq:abun}
\end{equation}

\noindent We measure a median carbon abundance of $\rm X[CI]/X[H_2]=3.9\pm0.4 \times 10^{-5}$ which is comparable to the abundance quoted in  \cite{Weiss05} ($\sim$ $\rm X[CI]/X[H_2]=5 \times 10^{-5}$) and derived by \cite{Danielson11} of $\rm X[CI]/X[H_2]=3.8\pm0.1 \times 10^{-5}$ for the lensed SMG, J213511, however it is higher than the value found in the Galaxy of $2.2 \times 10^{-5}$ \citep{Frerking89}.

To calculate the molecular gas mass using the [C{\sc i}] line we then assume $Q_{10}$=0.49$\pm$0.02, the median  $Q_{10}$ derived by \cite{Papadopoulos04b} and use:

\begin{eqnarray}
\nonumber
\rm M_{[CI]}(H_2)= 1375.8\frac{D_l^2}{1+z}(\frac{X_{[CI]}}{10^{-5}})^{-1} \\
\times \rm (\frac{A_{10}}{10^{-7}s^{-1}})^{-1} Q_{10}^{-1} \frac{S_{[CI]}}{Jy\,km\,s^{-1}} [M_{\odot}]\\ \nonumber
\label{eq:massP}
\end{eqnarray}

\noindent  \citep{Wagg06,Papadopoulos04b}. Here $\rm X_{[CI]}=3\times 10^{-5}$  \citep{Weiss05} and $\rm A_{10}=7.93\times 10^{-8} s^{-1}$ (the Einstein A coefficient) and the cosmological dependence is encapsulated in D$_l$, the luminosity distance. We measure a median molecular gas mass from [C{\sc i}](1--0) of  $\rm 80\pm20 \times 10^9M_{\odot}$ which is consistent with the measure from the $^{12}$CO(1--0) luminosity, indicating that the [C{\sc i}](1--0) line luminosity is a viable tracer of the H$_2$ mass, regardless of our inconclusive findings from the line width comparisons in Section \ref{sec:profile}.  We note, however, that a $\rm X_{[CI]}$ must still be assumed, which is also a weakness of using any optically thin tracer of H$_2$ gas mass, such as $^{13}$CO lines. In this context the [C{\sc i}] lines with their simpler 3-level partition function, and simpler chemistry retain an advantage over the $^{13}$CO lines. On the other hand, using optically thick $^{12}$CO lines does not directly suffer from the corresponding $^{12}$CO/H$_2$ abundance uncertainty, but from other, different, uncertainties that we discuss in Section \ref{sec:phi}. However in  cases where direct measurements of $^{12}$CO(1--0) are not available or possible,  the [C{\sc i}](1--0) luminosity appears to be a reasonable substitute to estimate the  H$_2$ gas mass.

We note that measuring [C{\sc i}](2--1), which is readily achievable with ALMA at these redshifts, will enable mapping of the gas properties while avoiding assumptions of the gas excitation temperature, determined from the ratio of [C{\sc i}](2--1)\,/\,[C{\sc i}](1--0). This is of particular importance for studying southern sources with ALMA where sensitive measurements at the observed frequencies of $\sim30$GHz for $^{12}$CO(1--0) are not  achievable, but the [C{\sc i}] lines are accessible.

\begin{table}
\centering
\begin{tabular}{|l|c|c|c|c|c|c|c|c|c|}
\hline\hline
ID & $\rm M_{[CI]}$ & $\rm M_{[CI]}(H_2)$ & $\rm M_{CO}(H_2)$\\
   & $\rm 10^7M_{\odot}$ & $\rm 10^9M_{\odot}$ & $\rm 10^{9}M_{\odot}$ \\
\hline\hline
SXDF7    & 1.6$\pm$0.4   & 120$\pm$30 & 70$\pm$20\\
SXDF11   & $<$1.2        & $<$90   & 39$\pm$8\\ 
SXDF4a   & 0.9$\pm$0.3   & 70$\pm$20   & 30$\pm$10\\
SXDF4b   & $<$1.0        &$<$70        & 50$\pm$20\\
SA22.96   & 1.5$\pm$0.5   & 110$\pm$30  & 60$\pm$20\\    
J02399   & 1.8$\pm$0.2   & 130$\pm$20  & 80$\pm$20\\
J123549  & 1.8$\pm$0.3   & 130$\pm$20     & 74$\pm$9\\
J163650  & $<$1.3        & $<$90      & 90$\pm$10\\
J163658  & 1.7$\pm$0.4   & 120$\pm$30     & 100$\pm$20\\
J14011   & 0.9$\pm$0.2   & 70$\pm$10 & 24$\pm$3\\
J16359   & 0.16$\pm$0.03 & 11$\pm$2 & 8.2$\pm$0.8\\
J213511  & 0.87$\pm$0.03 & 61$\pm$3 & 17.1$\pm$0.9\\
ID141    & 0.7$\pm$0.2   & 50$\pm$20 & 39$\pm$5\\ 
MM18423+5938 & 0.5$\pm$0.1 & 34$\pm$8 & 23.1$\pm$0.2\\
\hline\hline
\end{tabular}
\caption{The neutral carbon masses and molecular gas masses calculated from the $^{12}$CO(1--0) and [C{\sc i}](1--0) luminosities for the SMGs with new [C{\sc i}](1--0) observations in this work and the literature sample of SMGs.}
\label{tab:mass}
\end{table}

\section{Discussion}
\label{sec:dis}

\subsection{Physical Interpretation}
\label{sec:phi}
We next explore what the derived conditions in the ISM might imply for  the SMG population.  We first note that the density and radiation field values derived from the comparison to the PDR models may not represent the true values for these properties and that there are a number of issues with PDR modelling as described in Section \ref{sec:pdrissues}.  Furthermore, since we only use a single crossing (the intersection of two line ratios only) to constrain the densities and radiation fields we are not sampling the model distribution well, as shown by the spread of values derived by \cite{Danielson11} using multiple line ratios. However, we use the single crossing to derive values in both the SMG and quasar sample, therefore performing a useful and consistent {\it differential} comparison between the two samples. 

We find that the SMGs are clearly separate from the quasars in terms of radiation field but have similar densities. We also find that 5 of the 14 SMGs (mostly from our new observations in this work) extend the range of SMG radiation fields to much lower values, almost ten times lower than the majority of previous SMGs studied.  This tail of SMGs' $G_0$ are more consistent with starbursts or normal, local star-forming galaxies than with ULIRGs. Such a scenario is often quoted in the recent SMG literature from a variety of other arguments, namely the large sizes of SMGs \citep[e.g.,]{Chapman04b,Bothwell10}, or the suppressed PAH line ratios \citep{Menendez-Delmestre09}.

In order to bring these low-$G_0$ SMGs to the  median of the literature SMG sample, the ratio of $ \rm L_{[CI](1-0)}/L_{FIR}$ would need to be lowered by a factor $\sim4\times$. However the  $\rm L_{FIR}$ values are well constrained since we use a combination of 2\,mm, 850\,mm, {\it Herschel}-SPIRE and radio fluxes to constrain the SEDs \citep{Alaghband-Zadeh12}, leaving less than a factor two in uncertainty. Therefore we suggest that the low radiation field values are more likely due to boosted [C{\sc i}] fluxes relative to the $\rm L_{FIR}$. Enhanced [C{\sc i}] is predicted by \cite{Papadopoulos04a} to occur in star forming environments since the processes which make [C{\sc i}] present throughout a typical giant molecular cloud act to enhance it further in a star forming or dynamically fast-evolving environment (for example, non-equilibrium chemical states and turbulent diffuse mixing). However, it is also possible that high densities (n$>$10$^4$, which is the case for the SMGs) that prevail in the ISM of merger and starburst systems may also counteract this boosting and suppress the [C{\sc i}]/H$_2$ abundance \citep{Papadopoulos04b} therefore there are a number of uncertainties in this prediction since the balance between the two effects is not well constrained. The PDR models also assume a certain geometry and a homogeneous medium \citep{Danielson11} which may result in lower predicted [C{\sc i}] in the models compared to our observations.

We note that the large $ \rm L_{[CI](1-0)}/L_{CO(4-3)}$ and $ \rm L_{[CI](1-0)}/L_{FIR}$ values (Fig. \ref{fig:prop3}), causing low G$_0$ values in some SMGs, may be a result of a more extended, cool  H$_2$ gas distribution. This ``non-star-forming gas'' would be weaker in the far-infrared and $^{12}$CO(4--3) emission but strong in [C{\sc i}].  Furthermore, if only a small fraction of the [C{\sc i}] emission comes from PDRs (Section \ref{sec:pdrissues}) then the PDR-only ratios of $ \rm L_{[CI](1-0)}/L_{CO(4-3)}$ and $ \rm L_{[CI](1-0)}/L_{FIR}$ in Fig. \ref{fig:prop3} should be lower since the non-PDR contribution to the [C{\sc i}] emission should be subtracted from the observed values. This would imply that the density and radiation field values should be higher still for the PDR-related gas.

In order to explore the extent of the star formation in the SMG sample we derive sizes of the star forming regions using the molecular gas masses inferred in Section \ref{sec:mass} and the densities determined from the PDR analysis in Section \ref{sec:pdr}.  We first note that the densities derived from the PDR analysis may not represent the true densities owing the large evidence that [C{\sc i}] does not have a PDR-like distribution, as discussed in detail in Section \ref{sec:pdrissues}. However since we are only considering the average densities of the whole galaxy systems to gain approximate size estimates we continue to use the PDR-derived densities. We use the H$_2$ masses derived from the $^{12}$CO(1--0) luminosities, applying a correction factor of 1.36$\times$ to account for the total mass of Helium and Hydrogen to gain the total molecular gas mass \citep{Tacconi10}. Using these corrected masses we determine a median diameter of the SMGs of 800$\pm$100\,pc assuming a geometry of a 500\,pc thick disc \citep{Lockman86} (800$\pm$80\,pc assuming spherical geometry). These derived diameters are smaller than we expect for SMGs where there is numerous evidence that the star formation on average is extended on scales greater than 1\,kpc. The extended star formation has been suggested through spectroscopy of PAH features, resolved radio continuum, and large resolved H$\alpha$ distributions -- \citealt{Menendez-Delmestre09, Chapman04b, Alaghband-Zadeh12}. Furthermore, \cite{Bothwell10} find extended $^{12}$CO structures in three SMGs, with a mean radius of 3.1\,kpc. Indeed for a number of the SMGs in our sample the sizes have been derived from high resolution $^{12}$CO and H$\alpha$ observations. We find that the sizes derived from the densities and masses in this work underestimate these sizes. \cite{Tacconi08} measure diameters of 1.9$\pm$0.8\,kpc and 2$\pm$1\,kpc for J123549 and J163658 respectively whereas we derive diameters of 0.9$\pm$0.1\,kpc and 1.0$\pm$0.2\,kpc. In \cite{Alaghband-Zadeh12} we derive H$\alpha$ sizes for SXDF7, SXDF4a and SA22.96 finding diameters of 5.2$\pm$0.1\,kpc, 5.8$\pm$0.1\,kpc and 8.4$\pm$0.6\,kpc respectively which are larger than the H$_2$ mass derived diameters of 1.0$\pm$0.1\,kpc, 0.7$\pm$0.1\,kpc and  1.0$\pm$0.2\,kpc. 

However, the SMGs often comprise multiple star-forming components with clumps of star formation across the complicated systems. Clear examples are demonstrated by the H$\alpha$ observations of SMGs in \cite{Alaghband-Zadeh12}, or the clumps in the lensed `Eyelash' SMG \citep{Swinbank10a}. The derived sizes are therefore minimum sizes since the effects of mass clumping would be to make the derived sizes larger. For spherical geometry the radius of the region is defined as:

\begin{equation} 
\rm r=(\frac{3M_{H_2}}{4\pi f_cn_c})^{1/3}
\end{equation}

\noindent  where $\rm f_c$ is the density clumping factor, equal to the ratio of the mean gas density (n) to the clump gas density yielded from line ratio models for the average molecular cloud ensemble ($\rm n_c$). For $\rm f_c<1$ the radii derived are therefore minimum values. Since the median  800$\pm$100\,pc size is a minimum, we find that the sizes are therefore likely consistent with being extended on kpc scales.

We also note that the derived H$_2$ masses from the $^{12}$CO(1--0) luminosities assume a conversion factor from $^{12}$CO(1--0) luminosity to H$_2$ mass of  $\rm \alpha_{CO}=1.0\,M_{\odot}(K km s^{-1} pc^2)^{-1}$. It is possible that this value may be larger in some star forming galaxies. There is evidence for $\rm \alpha_{CO} \sim 0.8-5\,M_{\odot} (K km s^{-1} pc^2)^{-1}$ in a range of star forming galaxies \citep{Tacconi08,Daddi10a,Ivison11,Hodge12}. Furthermore, there is evidence that  $\rm \alpha_{CO}$ can be as high as $\rm \sim3-6\,M_{\odot}(K km s^{-1} pc^2)^{-1}$ (near Galactic) in local ULIRGs from the work of \cite{Papadopoulos12b}. We note that in this case (of $\rm \alpha_{CO} \sim 6\,M_{\odot}(K km s^{-1} pc^2)^{-1}$) the [C{\sc i}]/H$_2$ abundance must be $\sim$6 times smaller in order for the gas masses derived to match the $^{12}$CO(1--0) derived gas masses. This may be achieved if the high average gas densities in the SMGs suppress the average  [C{\sc i}]/H$_2$ abundance \citep{Papadopoulos04b} while the same high gas densities boost $\rm \alpha_{CO}$. \cite{Papadopoulos12b} suggest that well-sampled high-J $^{12}$CO lines in combination with heavy rotor molecules such as HCN are required to constrain the $\rm \alpha_{CO}$ in ULIRGs where the turbulent gas, with a strong high-density component, will have a high $\rm \alpha_{CO}$ which dominates in the system. There are further uncertainties in the values of $\rm \alpha_{CO}$ in different environments as detailed by \cite{Bolatto13} such as a dependence of  $\rm \alpha_{CO}$ on metallicity and large differences within galaxies from the centres to the outer disk regions. 

Furthermore, the majority of SMGs are merger systems \citep{Engel10,Alaghband-Zadeh12} and we note that torque induced by a merger event creates a more centralised, homogeneous molecular ISM, in which pressure plays a significant role in triggering a star forming event as opposed to in normal star formation which occurs in gravitationally bound molecular clouds (which exist as well defined separate structures in the ISM). It is therefore possible that in these merger systems the ISM could be approaching a high pressure, continuous medium, rather than the well separated giant molecular clouds found in a lower density starburst environment. In this case there is evidence that a different $\alpha_{CO}$ is required \citep{Ivison11}.  There are therefore many issues to consider when choosing the value of $\rm \alpha_{CO}$ to use, with further modelling and observations required to constrain the value across a range of environments.

We find that by adopting a conversion factor in the middle of the ranges discussed above ($\rm \alpha_{CO}=3\,M_{\odot} (K km s^{-1} pc^2)^{-1}$) the sizes we derive increase to $>$1\,kpc which is already consistent with what we expect and provides further support for extended star formation occurring in SMGs.

\subsection{$^{12}$CO(4--3)/[CI](1--0) as an indicator of Mergers versus Disks}
Finally, we explore whether the $^{12}$CO(4--3)/[C{\sc i}](1--0) ratio might act as a useful indicator of major mergers in SMGs. The extended star formation suggested by the large derived sizes in the SMGs (assuming a mid-range $\rm \alpha_{CO}=3\,M_{\odot} (K km s^{-1} pc^2)^{-1}$) may provide further support for the hypothesis that SMGs are merger systems \citep{Engel10,Alaghband-Zadeh12} since the star formation which occurs outside the central nuclei and across the systems  may be the relics of the progenitor galaxies of the merger, suggesting that we are observing multiple components in the early stages of a merger.  This implies that the ultra-luminous phase occurring in these systems may be triggered at the first stages of a merger, a possibility suggested in the hydrodynamical models of \cite{Narayanan09}, proving a valuable insight into the merger and star formation trigger processes.  We note that this contrasts with observations of merger-induced local ULIRGs which are found to be compact \citep{Sanders88, Sakamoto08} offering a difference between the ULIRG triggers at low and high redshift. We require high-resolution observations of the $^{12}$CO and [C{\sc i}] transitions to explore this further.

\cite{Papadopoulos12} probe the use of the $^{12}$CO(4--3) to [C{\sc i}](1--0) flux ratio as an indicator of the star formation mode in a system, suggesting whether a system is merger-driven or disk-like. They find that there is a difference of up to a factor of 10 between these two types of systems from a compilation of literature values of the ratio. They obtain $\rm \langle r_{CO(4-3)/[CI](1-0)}\rangle =4.6\pm1.5$ for merger-driven starbursts in ULIRGs and $\rm \langle r_{CO(4-3)/[C{\sc i}]}\rangle=0.45-1.3$ in disk-dominated star forming systems. We measure a median $\rm L'_{CO(4-3)}/L'_{[CI](1-0)}$ of $2.5\pm0.2$ providing evidence that the SMGs are merger-driven systems as opposed to quiescent, disk-like systems.

\section{Conclusions} 
\label{sec:conc}
New [C{\sc i}](1--0) and $^{12}$CO(4--3) observations of five SMGs are presented and combined with literature SMG [C{\sc i}] detections to probe the gas distribution and properties within these massive merging systems. We compare the measured line luminosity ratios of [C{\sc i}](1--0), various $^{12}$CO transitions and the far-infrared luminosity to the outputs of PDR models to constrain the physical conditions in the ISM of these systems, finding that the SMGs have densities and radiation field values that are consistent with local ULIRGs with 5 of the sample of 14 extending to lower densities and weaker radiation fields, more similar to the local starbursts, providing further support that star formation in SMGs can be extended. We find a sample of quasars have higher densities and radiation fields on average than the SMGs, consistent with the more extreme local ULIRGs, and reinforcing their interpretation as transition objects. Besides the standard interpretation of line ratios such as [C{\sc i}](1--0)/$^{12}$CO(4--3) in terms of PDRs and the PDR-model link to the average far-UV fields we also consider a non-PDR origin of the [C{\sc i}] line emission, with much of it actually tracing the bulk H$_2$ gas rather than the  H$_2$ gas only near star formation sites. We therefore explore the use of [C{\sc i}] as a tracer of the H$_2$ gas finding that it provides a mass that is consistent with the H$_2$ gas mass determined from the $^{12}$CO observations, providing an alternative measure of the H$_2$ gas in high-redshift systems without $^{12}$CO observations. [C{\sc i}] therefore provides an H$_2$ tracer which is of great use for high-redshift objects in which the low-J $^{12}$CO lines are redshifted to frequencies lower than the ALMA operating frequencies.  Owing to the likely clumping in SMGs and the low value of the conversion between  $^{12}$CO luminosity and H$_2$ gas mass used, the SMG sizes we derive are minima. We find that the minimum sizes are likely consistent with the sizes derived from high-resolution $^{12}$CO and H$\alpha$ observations, providing further support for extended star formation in SMGs. Finally, the uncertainties in the excitation temperature highlight the need for complementary [C{\sc i}](2--1) observations in these SMGs to enable [C{\sc i}] to be used as a reliable tracer of molecular gas in high-z galaxies and therefore to probe galaxy evolution out to the epoch of the peak of star formation in the Universe.

\section{Acknowledgements}
\label{sec:ackn}
This paper is based on observations made at IRAM- PdBI under program number V040. SA-Z acknowledges the support from STFC. IRS acknowledges the support from the Leverhulme Trust and STFC.
\bibliographystyle{mn2e}

\bibliography{refs_short}

\begin{thebibliography}{84}
\expandafter\ifx\csname natexlab\endcsname\relax\def\natexlab#1{#1}\fi

\bibitem[{{Alaghband-Zadeh} {et~al.}(2012)}]{Alaghband-Zadeh12}
{Alaghband-Zadeh} S., {et~al.}, 2012, MNRAS, 424, 2232

\bibitem[{{Ao} {et~al.}(2008){Ao}, {Wei{\ss}}, {Downes}, {Walter}, {Henkel}, \&
  {Menten}}]{Ao08}
{Ao} Y., {Wei{\ss}} A., {Downes} D., {Walter} F., {Henkel} C., {Menten} K.~M.,
  2008, A\&A, 491, 747

\bibitem[{{Barvainis} {et~al.}(1997){Barvainis}, {Maloney}, {Antonucci}, \&
  {Alloin}}]{Barvainis97}
{Barvainis} R., {Maloney} P., {Antonucci} R., {Alloin} D., 1997, ApJ, 484, 695

\bibitem[{{Biggs} \& {Ivison}(2008)}]{Biggs08}
{Biggs} A.~D., {Ivison} R.~J., 2008, MNRAS, 385, 893

\bibitem[{{Bland} \& {Tully}(1988)}]{Bland88}
{Bland} J., {Tully} B., 1988, Nature, 334, 43

\bibitem[{{Bolatto} {et~al.}(2013){Bolatto}, {Wolfire}, \& {Leroy}}]{Bolatto13}
{Bolatto} A.~D., {Wolfire} M., {Leroy} A.~K., 2013, arXiv:1301.3498

\bibitem[{{Bothwell} {et~al.}(2010)}]{Bothwell10}
{Bothwell} M.~S., {et~al.}, 2010, MNRAS, 405, 219

\bibitem[{{Bothwell} {et~al.}(2013)}]{Bothwell12}
---, 2013, MNRAS, 429, 3047

\bibitem[{{Bradford} {et~al.}(2003)}]{Bradford03}
{Bradford} C.~M., {et~al.}, 2003, ApJ, 586, 891

\bibitem[{{Chapman} {et~al.}(2005){Chapman}, {Blain}, {Smail}, \&
  {Ivison}}]{Chapman05}
{Chapman} S.~C., {Blain} A.~W., {Smail} I., {Ivison} R.~J., 2005, ApJ, 622, 772

\bibitem[{{Chapman} {et~al.}(2004){Chapman}, {Smail}, {Windhorst}, {Muxlow}, \&
  {Ivison}}]{Chapman04b}
{Chapman} S.~C., {Smail} I., {Windhorst} R., {Muxlow} T., {Ivison} R.~J., 2004,
  ApJ, 611, 732

\bibitem[{{Coppin} {et~al.}(2006)}]{Coppin06}
{Coppin} K., {et~al.}, 2006, MNRAS, 372, 1621

\bibitem[{{Cox} {et~al.}(2011)}]{Cox11}
{Cox} P., {et~al.}, 2011, ApJ, 740, 63

\bibitem[{{Daddi} {et~al.}(2010)}]{Daddi10a}
{Daddi} E., {et~al.}, 2010, ApJ, 713, 686

\bibitem[{{Danielson} {et~al.}(2011)}]{Danielson11}
{Danielson} A.~L.~R., {et~al.}, 2011, MNRAS, 410, 1687

\bibitem[{{Davies} {et~al.}(2003){Davies}, {Sternberg}, {Lehnert}, \&
  {Tacconi-Garman}}]{Davies03}
{Davies} R.~I., {Sternberg} A., {Lehnert} M., {Tacconi-Garman} L.~E., 2003,
  ApJ, 597, 907

\bibitem[{{Downes} \& {Solomon}(2003)}]{Downes03}
{Downes} D., {Solomon} P.~M., 2003, ApJ, 582, 37

\bibitem[{{Engel} {et~al.}(2010)}]{Engel10}
{Engel} H., {et~al.}, 2010, ApJ, 724, 233

\bibitem[{{Fixsen} {et~al.}(1999){Fixsen}, {Bennett}, \& {Mather}}]{Fixsen99}
{Fixsen} D.~J., {Bennett} C.~L., {Mather} J.~C., 1999, ApJ, 526, 207

\bibitem[{{Frerking} {et~al.}(1989){Frerking}, {Keene}, {Blake}, \&
  {Phillips}}]{Frerking89}
{Frerking} M.~A., {Keene} J., {Blake} G.~A., {Phillips} T.~G., 1989, ApJ, 344,
  311

\bibitem[{{Genzel} {et~al.}(2003)}]{Genzel03}
{Genzel} R., {et~al.}, 2003, ApJL, 584, 633

\bibitem[{{Gerin} \& {Phillips}(1998)}]{Gerin98}
{Gerin} M., {Phillips} T.~G., 1998, ApJL, 509, L17

\bibitem[{{Gerin} \& {Phillips}(2000)}]{Gerin00}
---, 2000, ApJ, 537, 644

\bibitem[{{Gerin} {et~al.}(2002){Gerin}, {Phillips}, \& {Contursi}}]{Gerin02}
{Gerin} M., {Phillips} T.~G., {Contursi} A., 2002, in EAS Publications Series,
  Vol.~4, EAS Publications Series, {Giard} M., {Bernard} J.~P., {Klotz} A.,
  {Ristorcelli} I., eds., pp. 193--193

\bibitem[{{Greve} {et~al.}(2005)}]{Greve05}
{Greve} T.~R., {et~al.}, 2005, MNRAS, 359, 1165

\bibitem[{{Hailey-Dunsheath} {et~al.}(2012)}]{Hailey-Dunsheath12}
{Hailey-Dunsheath} S., {et~al.}, 2012, ApJ, 755, 57

\bibitem[{{Hainline} {et~al.}(2009){Hainline}, {Blain}, {Smail}, {Frayer},
  {Chapman}, {Ivison}, \& {Alexander}}]{Hainline09}
{Hainline} L.~J., {Blain} A.~W., {Smail} I., {Frayer} D.~T., {Chapman} S.~C.,
  {Ivison} R.~J., {Alexander} D.~M., 2009, ApJ, 699, 1610

\bibitem[{{Hodge} {et~al.}(2012)}]{Hodge12}
{Hodge} J.~A., {et~al.}, 2012, ApJ, 760, 11

\bibitem[{{Hollenbach} {et~al.}(1991){Hollenbach}, {Takahashi}, \&
  {Tielens}}]{Hollenbach91}
{Hollenbach} D.~J., {Takahashi} T., {Tielens} A.~G.~G.~M., 1991, ApJ, 377, 192

\bibitem[{{Hollenbach} \& {Tielens}(1999)}]{Hollenbach99}
{Hollenbach} D.~J., {Tielens} A.~G.~G.~M., 1999, Reviews of Modern Physics, 71,
  173

\bibitem[{{Ikeda} {et~al.}(2002){Ikeda}, {Oka}, {Tatematsu}, {Sekimoto}, \&
  {Yamamoto}}]{Ikeda02}
{Ikeda} M., {Oka} T., {Tatematsu} K., {Sekimoto} Y., {Yamamoto} S., 2002, ApJS,
  139, 467

\bibitem[{{Israel} \& {Baas}(2002)}]{Israel02}
{Israel} F.~P., {Baas} F., 2002, A\&A, 383, 82

\bibitem[{{Ivison} {et~al.}(2010)}]{Ivison10b}
{Ivison} R.~J., {et~al.}, 2010, A\&A, 518, L35

\bibitem[{{Ivison} {et~al.}(2011)}]{Ivison11}
---, 2011, MNRAS, 412, 1913

\bibitem[{{Jansen} {et~al.}(1995){Jansen}, {van Dishoeck}, {Black}, {Spaans},
  \& {Sosin}}]{Jansen95}
{Jansen} D.~J., {van Dishoeck} E.~F., {Black} J.~H., {Spaans} M., {Sosin} C.,
  1995, A\&A, 302, 223

\bibitem[{{Kaufman} {et~al.}(1999){Kaufman}, {Wolfire}, {Hollenbach}, \&
  {Luhman}}]{Kaufman99}
{Kaufman} M.~J., {Wolfire} M.~G., {Hollenbach} D.~J., {Luhman} M.~L., 1999,
  ApJ, 527, 795

\bibitem[{{Keene} {et~al.}(1985){Keene}, {Blake}, {Phillips}, {Huggins}, \&
  {Beichman}}]{Keene85}
{Keene} J., {Blake} G.~A., {Phillips} T.~G., {Huggins} P.~J., {Beichman} C.~A.,
  1985, ApJ, 299, 967

\bibitem[{{Keene} {et~al.}(1996){Keene}, {Lis}, {Phillips}, \&
  {Schilke}}]{Keene96}
{Keene} J., {Lis} D.~C., {Phillips} T.~G., {Schilke} P., 1996, in IAU
  Symposium, Vol. 178, Molecules in Astrophysics: Probes \& Processes, {van
  Dishoeck} E.~F., ed., p. 129

\bibitem[{{Lestrade} {et~al.}(2010){Lestrade}, {Combes}, {Salom{\'e}}, {Omont},
  {Bertoldi}, {Andr{\'e}}, \& {Schneider}}]{Lestrade10}
{Lestrade} J.-F., {Combes} F., {Salom{\'e}} P., {Omont} A., {Bertoldi} F.,
  {Andr{\'e}} P., {Schneider} N., 2010, A\&A, 522, L4

\bibitem[{{Lockman} {et~al.}(1986){Lockman}, {Hobbs}, \& {Shull}}]{Lockman86}
{Lockman} F.~J., {Hobbs} L.~M., {Shull} J.~M., 1986, ApJ, 301, 380

\bibitem[{{Magnelli} {et~al.}(2012)}]{Magnelli12}
{Magnelli} B., {et~al.}, 2012, A\&A, 539, A155

\bibitem[{{Maiolino} {et~al.}(2005)}]{Maiolino05}
{Maiolino} R., {et~al.}, 2005, A\&A, 440, L51

\bibitem[{{Malhotra} {et~al.}(2001)}]{Malhotra01}
{Malhotra} S., {et~al.}, 2001, ApJ, 561, 766

\bibitem[{{Meijerink} {et~al.}(2013)}]{Meijerink13}
{Meijerink} R., {et~al.}, 2013, ApJL, 762, L16

\bibitem[{{Meixner} \& {Tielens}(1993)}]{Meixner93}
{Meixner} M., {Tielens} A.~G.~G.~M., 1993, ApJ, 405, 216

\bibitem[{{Men{\'e}ndez-Delmestre} {et~al.}(2009)}]{Menendez-Delmestre09}
{Men{\'e}ndez-Delmestre} K., {et~al.}, 2009, ApJ, 699, 667

\bibitem[{{Narayanan} {et~al.}(2009){Narayanan}, {Cox}, {Hayward}, {Younger},
  \& {Hernquist}}]{Narayanan09}
{Narayanan} D., {Cox} T.~J., {Hayward} C.~C., {Younger} J.~D., {Hernquist} L.,
  2009, MNRAS, 400, 1919

\bibitem[{{Ojha} {et~al.}(2001)}]{Ojha01}
{Ojha} R., {et~al.}, 2001, ApJ, 548, 253

\bibitem[{{Oka} {et~al.}(2001)}]{Oka01}
{Oka} T., {et~al.}, 2001, ApJ, 558, 176

\bibitem[{{Papadopoulos} \& {Geach}(2012)}]{Papadopoulos12}
{Papadopoulos} P.~P., {Geach} J.~E., 2012, ApJ, 757, 157

\bibitem[{{Papadopoulos} \& {Greve}(2004)}]{Papadopoulos04b}
{Papadopoulos} P.~P., {Greve} T.~R., 2004, ApJL, 615, L29

\bibitem[{{Papadopoulos} {et~al.}(2004){Papadopoulos}, {Thi}, \&
  {Viti}}]{Papadopoulos04a}
{Papadopoulos} P.~P., {Thi} W.-F., {Viti} S., 2004, MNRAS, 351, 147

\bibitem[{{Papadopoulos} {et~al.}(2012){Papadopoulos}, {van der Werf},
  {Xilouris}, {Isaak}, {Gao}, \& {M{\"u}hle}}]{Papadopoulos12b}
{Papadopoulos} P.~P., {van der Werf} P.~P., {Xilouris} E.~M., {Isaak} K.~G.,
  {Gao} Y., {M{\"u}hle} S., 2012, MNRAS, 426, 2601

\bibitem[{{Pety} {et~al.}(2004){Pety}, {Beelen}, {Cox}, {Downes}, {Omont},
  {Bertoldi}, \& {Carilli}}]{Pety04}
{Pety} J., {Beelen} A., {Cox} P., {Downes} D., {Omont} A., {Bertoldi} F.,
  {Carilli} C.~L., 2004, A\&A, 428, L21

\bibitem[{{Phillips} \& {Huggins}(1981)}]{Phillips81}
{Phillips} T.~G., {Huggins} P.~J., 1981, ApJ, 251, 533

\bibitem[{{Pineda} \& {Bensch}(2007)}]{Pineda07}
{Pineda} J.~L., {Bensch} F., 2007, A\&A, 470, 615

\bibitem[{{Plume}(1995)}]{Plume95}
{Plume} R., 1995, PhD thesis, THE UNIVERSITY OF TEXAS AT AUSTIN.

\bibitem[{{Rangwala} {et~al.}(2011)}]{Rangwala11}
{Rangwala} N., {et~al.}, 2011, ApJ, 743, 94

\bibitem[{{Riechers} {et~al.}(2009)}]{Riechers09}
{Riechers} D.~A., {et~al.}, 2009, ApJ, 703, 1338

\bibitem[{{R{\"o}llig} {et~al.}(2007)}]{Rollig07}
{R{\"o}llig} M., {et~al.}, 2007, A\&A, 467, 187

\bibitem[{{Sakamoto} {et~al.}(2008)}]{Sakamoto08}
{Sakamoto} K., {et~al.}, 2008, ApJ, 684, 957

\bibitem[{{Sanders} \& {Mirabel}(1996)}]{Sanders96}
{Sanders} D.~B., {Mirabel} I.~F., 1996, ARA\&A, 34, 749

\bibitem[{{Sanders} {et~al.}(1988)}]{Sanders88}
{Sanders} D.~B., {et~al.}, 1988, ApJ, 325, 74

\bibitem[{{Schneider} {et~al.}(2003){Schneider}, {Simon}, {Kramer}, {Kraemer},
  {Stutzki}, \& {Mookerjea}}]{Schneider03}
{Schneider} N., {Simon} R., {Kramer} C., {Kraemer} K., {Stutzki} J.,
  {Mookerjea} B., 2003, A\&A, 406, 915

\bibitem[{{Sharon} {et~al.}(2013){Sharon}, {Baker}, {Harris}, \&
  {Thomson}}]{Sharon12}
{Sharon} C.~E., {Baker} A.~J., {Harris} A.~I., {Thomson} A.~P., 2013, ApJ, 765,
  6

\bibitem[{{Smail} {et~al.}(2005){Smail}, {Smith}, \& {Ivison}}]{Smail05}
{Smail} I., {Smith} G.~P., {Ivison} R.~J., 2005, ApJ, 631, 121

\bibitem[{{Solomon} \& {Vanden Bout}(2005)}]{Solomon05}
{Solomon} P.~M., {Vanden Bout} P.~A., 2005, ARA\&A, 43, 677

\bibitem[{{Spaans}(1996)}]{Spaans96}
{Spaans} M., 1996, A\&A, 307, 271

\bibitem[{{Stacey} {et~al.}(1991){Stacey}, {Geis}, {Genzel}, {Lugten},
  {Poglitsch}, {Sternberg}, \& {Townes}}]{Stacey91}
{Stacey} G.~J., {Geis} N., {Genzel} R., {Lugten} J.~B., {Poglitsch} A.,
  {Sternberg} A., {Townes} C.~H., 1991, ApJ, 373, 423

\bibitem[{{Stoerzer} {et~al.}(1996){Stoerzer}, {Stutzki}, \&
  {Sternberg}}]{Storzer96}
{Stoerzer} H., {Stutzki} J., {Sternberg} A., 1996, A\&A, 310, 592

\bibitem[{{Stutzki} {et~al.}(1988)}]{Stutzki88}
{Stutzki} J., {et~al.}, 1988, ApJ, 332, 379

\bibitem[{{Swinbank} {et~al.}(2010)}]{Swinbank10a}
{Swinbank} A.~M., {et~al.}, 2010, Nature, 464, 733

\bibitem[{{Tacconi} {et~al.}(2008)}]{Tacconi08}
{Tacconi} L.~J., {et~al.}, 2008, ApJ, 680, 246

\bibitem[{{Tacconi} {et~al.}(2010)}]{Tacconi10}
---, 2010, Nature, 463, 781

\bibitem[{{Takata} {et~al.}(2006){Takata}, {Sekiguchi}, {Smail}, {Chapman},
  {Geach}, {Swinbank}, {Blain}, \& {Ivison}}]{Takata06}
{Takata} T., {Sekiguchi} K., {Smail} I., {Chapman} S.~C., {Geach} J.~E.,
  {Swinbank} A.~M., {Blain} A., {Ivison} R.~J., 2006, ApJ, 651, 713

\bibitem[{{Tatematsu} {et~al.}(1999){Tatematsu}, {Jaffe}, {Plume}, {Evans}, \&
  {Keene}}]{Tatematsu99}
{Tatematsu} K., {Jaffe} D.~T., {Plume} R., {Evans} II N.~J., {Keene} J., 1999,
  ApJ, 526, 295

\bibitem[{{Thomson} {et~al.}(2012)}]{Thomson12}
{Thomson} A.~P., {et~al.}, 2012, MNRAS, 425, 2203

\bibitem[{{Tielens} \& {Hollenbach}(1985)}]{Tielens85}
{Tielens} A.~G.~G.~M., {Hollenbach} D., 1985, ApJ, 291, 722

\bibitem[{{Van der Werf} {et~al.}(2010)}]{VanderWerf10}
{Van der Werf} P.~P., {et~al.}, 2010, A\&A, 518, L42

\bibitem[{{Wagg} {et~al.}(2006){Wagg}, {Wilner}, {Neri}, {Downes}, \&
  {Wiklind}}]{Wagg06}
{Wagg} J., {Wilner} D.~J., {Neri} R., {Downes} D., {Wiklind} T., 2006, ApJ,
  651, 46

\bibitem[{{Walter} {et~al.}(2011){Walter}, {Wei{\ss}}, {Downes}, {Decarli}, \&
  {Henkel}}]{Walter11}
{Walter} F., {Wei{\ss}} A., {Downes} D., {Decarli} R., {Henkel} C., 2011, ApJ,
  730, 18

\bibitem[{{Wardlow} {et~al.}(2011)}]{Wardlow11}
{Wardlow} J.~L., {et~al.}, 2011, MNRAS, 415, 1479

\bibitem[{{Wei{\ss}} {et~al.}(2005){Wei{\ss}}, {Downes}, {Henkel}, \&
  {Walter}}]{Weiss05}
{Wei{\ss}} A., {Downes} D., {Henkel} C., {Walter} F., 2005, A\&A, 429, L25

\bibitem[{{Wei{\ss}} {et~al.}(2003){Wei{\ss}}, {Henkel}, {Downes}, \&
  {Walter}}]{Weiss03}
{Wei{\ss}} A., {Henkel} C., {Downes} D., {Walter} F., 2003, A\&A, 409, L41

\end{thebibliography}

\appendix
\section{Supporting Figures - Constraining densities and radiation fields with molecular cloud models}

\begin{figure*}
\includegraphics[width=0.95\textwidth]{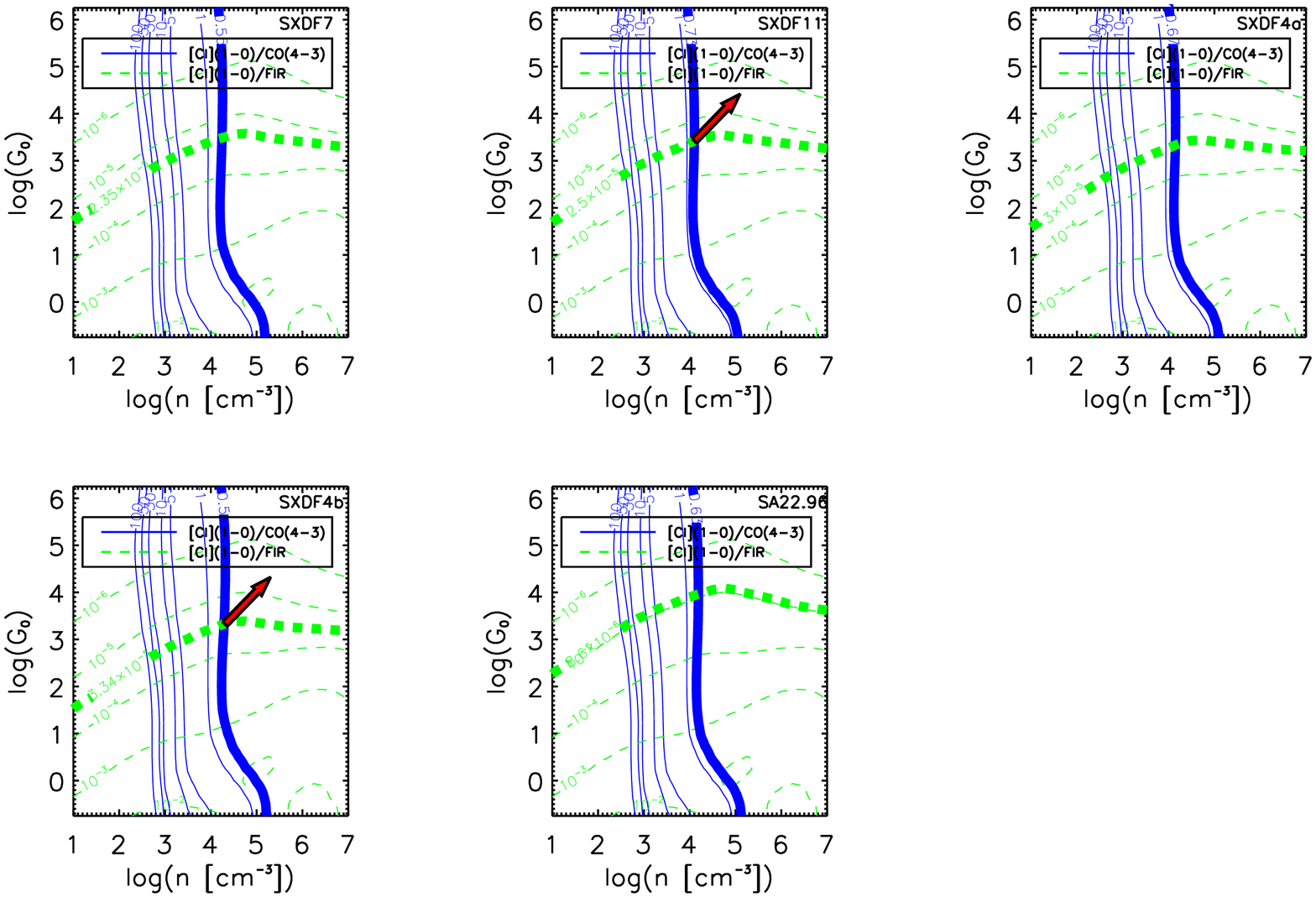}
\caption{Contours of $ \rm L_{[CI](1-0)}/L_{CO(4-3)}$ (solid blue)  and  $\rm L_{[CI](1-0)}/L_{FIR}$ (dashed green) at various levels of gas density ($n$) and radiation field ($G_0$) from the Photo-Dissociation Region (PDR) models of \protect\cite{Kaufman99}. The bold contours represent the observed ratios for each of the SMGs with new [C{\sc i}] detections presented in this work. The intersection of the two contours in each case provides a value for $n$ and $G_0$ for the SMG system. The red arrows mark the sources where there is only a limit to the [C{\sc i}](1--0) luminosity and therefore we can only gain a lower limit to $n$ and $G_0$.}
\label{fig:model}
\end{figure*}

\begin{figure*}
\includegraphics[width=0.95\textwidth]{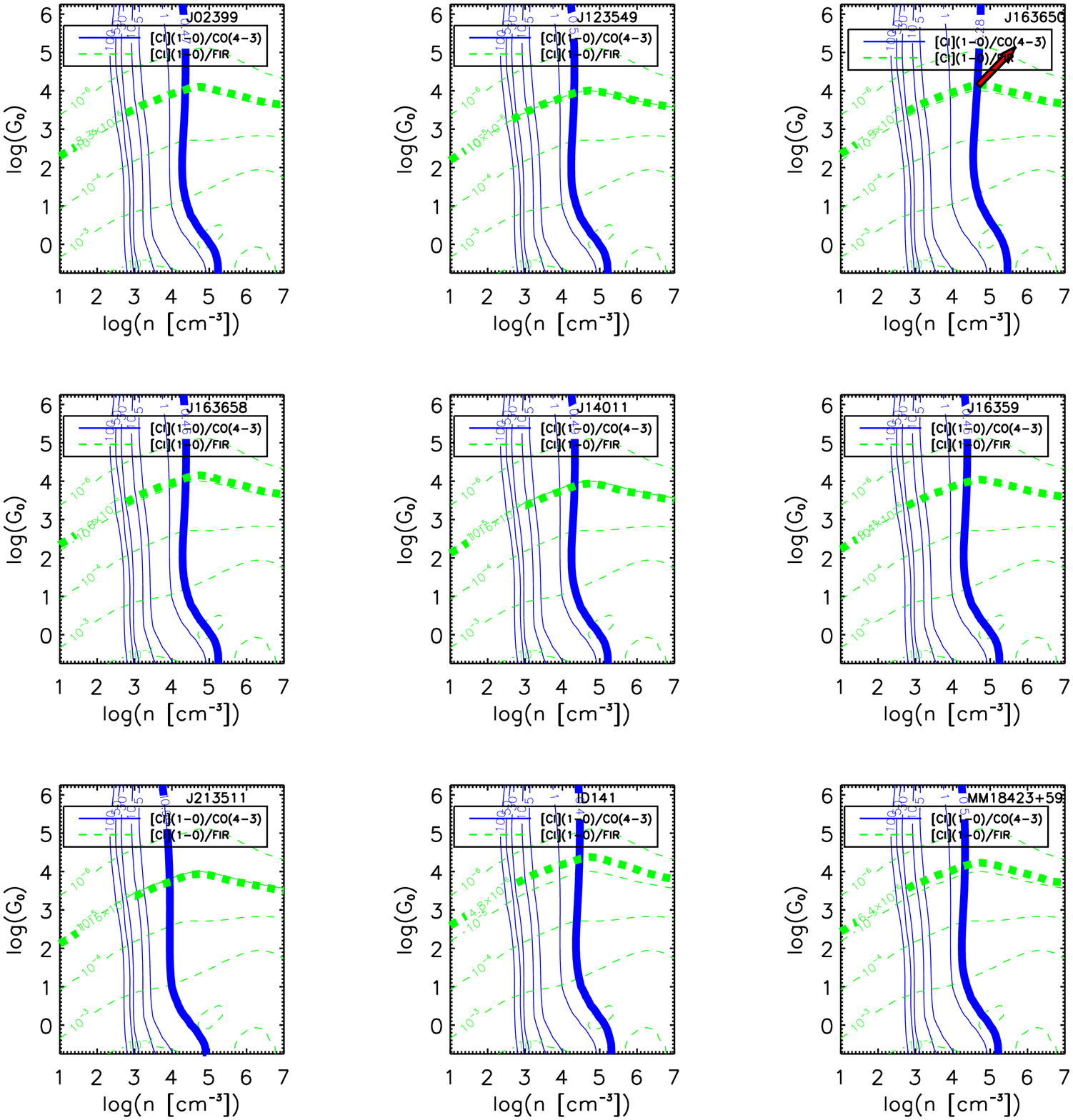}
\caption{As Fig. \protect\ref{fig:model} for the literature SMG sample. }
\label{fig:model_lit}
\end{figure*}

\begin{figure*}
\includegraphics[width=0.95\textwidth]{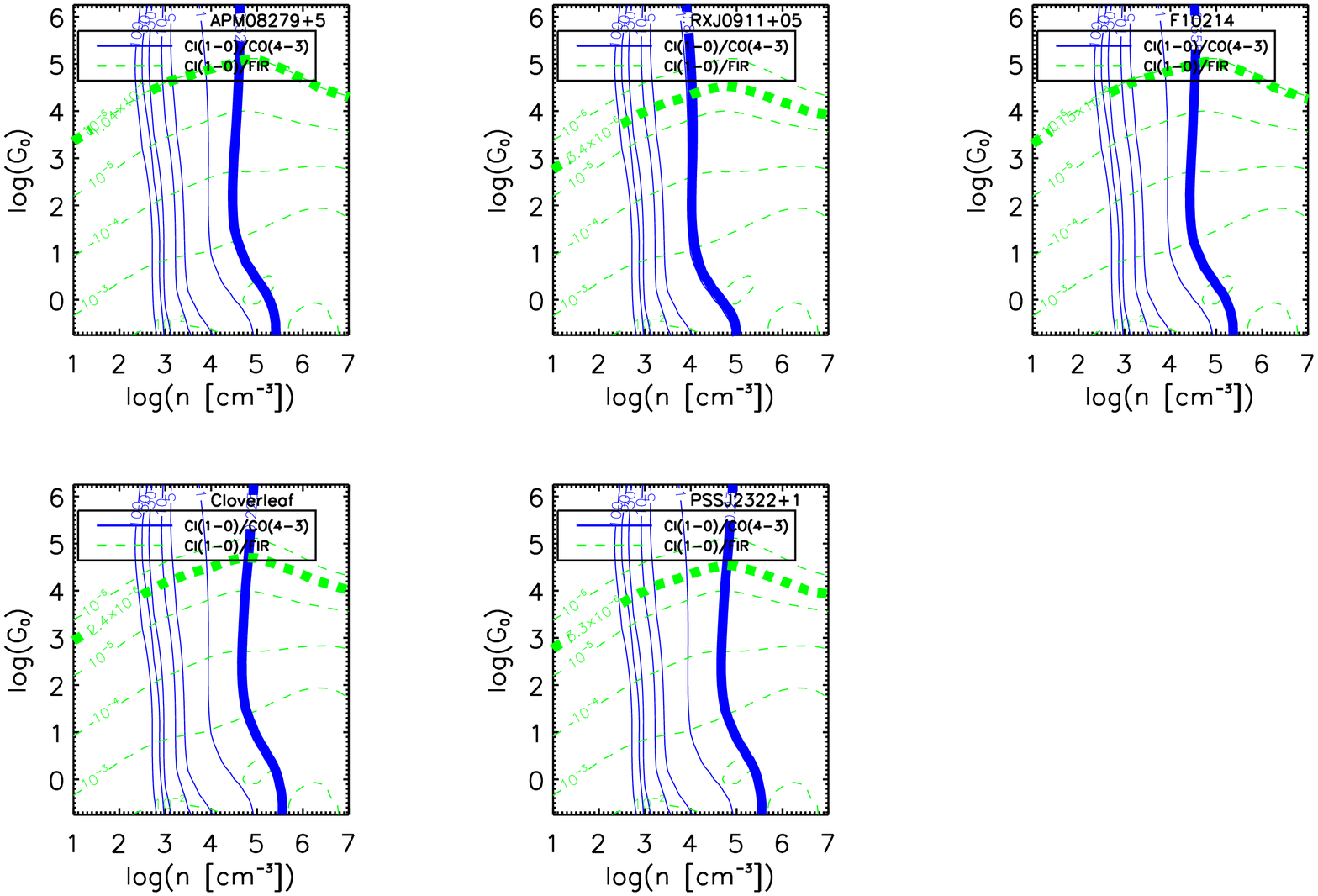}
\caption{As Fig. \protect\ref{fig:model} for the literature quasar sample.}
\label{fig:model_lit_nonsmg}
\end{figure*}

\end{document}